\documentclass{article}

\usepackage{float}

\usepackage{arxiv}
\usepackage{subcaption}
\usepackage{multirow}
\usepackage{amsmath}
\usepackage[utf8]{inputenc} % allow utf-8 input
\usepackage[T1]{fontenc}    % use 8-bit T1 fonts
\usepackage{hyperref}       % hyperlinks
\usepackage{url}            % simple URL typesetting
\usepackage{booktabs}       % professional-quality tables
\usepackage{amsfonts}       % blackboard math symbols
\usepackage{nicefrac}       % compact symbols for 1/2, etc.
\usepackage{microtype}      % microtypography
\usepackage{lipsum}
\usepackage{graphicx}
\graphicspath{ {./images/} }
\title{Well log data generation and imputation using sequence-based generative adversarial networks}

\author{
 Abdulrahman Al-Fakih* \\
  College of Petroleum Engineering and Geosciences\\
  King Fahd University of Petroleum \& Minerals\\
  Dhahran 31261, Saudi Arabia \\
  \texttt{alja2014ser@gmail.com} \\
   \And
   A. Koeshidayatullah* \\
  College of Petroleum Engineering and Geosciences\\
  King Fahd University of Petroleum \& Minerals\\
  Dhahran 31261, Saudi Arabia \\
  \texttt{a.koeshidayatullah@kfupm.edu.sa} \\
  \And
 Tapan Mukerji \\
  Departments of Energy Science \& Engineering,\\ Earth \& Planetary Sciences, and Geophysics\\
  University, Stanford, CA, USA\\
  \texttt{mukerji@stanford.edu} \\
  \And
    Sadam Al-Azani \\
  SDAIA-KFUPM Joint Research Center for Artificial Intelligence\\
  King Fahd University of Petroleum \& Minerals\\
  Dhahran 31261, Saudi Arabia \\
  \texttt{sadam.azani@kfupm.edu.sa} \\
  \And
SanLinn I. Kaka \\
  College of Petroleum Engineering and Geosciences\\
  King Fahd University of Petroleum \& Minerals\\
  Dhahran 31261, Saudi Arabia \\
  \texttt{skaka@kfupm.edu.sa} \\
}

\begin{document}
\maketitle
\begin{abstract}
Well log analysis is crucial for hydrocarbon exploration, providing detailed insights into subsurface geological formations. However, gaps and inaccuracies in well log data, often due to equipment limitations, operational challenges, and harsh subsurface conditions, can introduce significant uncertainties in reservoir evaluation. Addressing these challenges requires effective methods for both synthetic data generation and precise imputation of missing data, ensuring data completeness and reliability.
This study introduces a novel framework utilizing sequence-based generative adversarial networks (GANs) specifically designed for well log data generation and imputation. The framework integrates two distinct sequence-based GAN models: Time Series GAN (TSGAN) for generating synthetic well log data and Sequence GAN (SeqGAN) for imputing missing data. Both models were tested on a dataset from the North Sea, Netherlands region, focusing on different sections of 5, 10, and 50 data points.
Experimental results demonstrate that this approach achieves superior accuracy in filling data gaps compared to other deep learning models for spatial series analysis. The method yielded R² values of 0.921, 0.899, and 0.594, with corresponding mean absolute percentage error (MAPE) values of 8.320, 0.005, and 151.154, and mean absolute error (MAE) values of 0.012, 0.005, and 0.032, respectively. These results set a new benchmark for data integrity and utility in geosciences, particularly in well log data analysis.
\end{abstract}

\section*{Keywords}
Generative adversarial networks models; Time series generative adversarial networks models; Sequence generative adversarial networks models; Well log data imputation; Synthetic well log data generation.
\section*{Nomenclature}

\begin{tabbing}
\hspace{4cm} \= \hspace{3cm} \= \kill
BCE \> \> Binary Cross-Entropy \\
BRNN \> \> Bidirectional Recurrent Neural Networks \\
BRITS \> \> Bidirectional Recurrent Imputation for Time Series \\
CGAN \> \> Conditional Generative Adversarial Networks \\
CSV \> \> Comma-Separated Values \\
D \> \> Discriminator \\
DL \> \> Deep Learning \\
DNN \> \> Deep Neural Network \\
DT \> \> Sonic Transit Time \\
DTree \> \> Decision Trees \\
EDA \> \> Exploratory Data Analysis \\
FCNN \> \> Fully Connected Neural Network \\
GAN \> \> Generative Adversarial Networks \\
GMM \> \> Gaussian Mixture Models \\
G \> \> Generator \\
GR \> \> Gamma Ray \\
GRU \> \> Gated Recurrent Unit \\
ILD \> \> Induction Log Deep \\
IF \> \> Isolation Forest \\
KL \> \> Kullback-Leibler Divergence \\
KS \> \> Kolmogorov-Smirnov Test \\
LAS \> \> Log ASCII Standard \\
LOF \> \> Local Outlier Factor \\
NAOMI \> \> Non-Autoregressive Multiresolution Imputation \\
MAPE \> \> Mean Absolute Percentage Error \\
MAE \> \> Mean Absolute Error \\
ML \> \> Machine Learning \\
MRLE \> \> Mean Relative Log Error \\
MSE \> \> Mean Squared Error \\
NPHI \> \> Neutron Porosity \\
PCA \> \> Principal Component Analysis \\
PCC \> \> Pearson Correlation Coefficient \\
PHI \> \> Porosity \\
PHIECAL \> \> Effective Porosity Calculated \\
R2 \> \> Coefficient of Determination \\
RHOB \> \> Bulk Density \\
RF \> \> Random Forest \\
RNNs \> \> Recurrent Neural Networks \\
SeqGAN \> \> Sequence Generative Adversarial Networks \\
SVM \> \> One-Class Support Vector Machine \\
t-SNE \> \> t-Distributed Stochastic Neighbor Embedding \\
TL \> \> Transfer Learning \\
TSGAN \> \> Time Series GAN \\
VSHALE \> \> Volumetric Shale \\
\end{tabbing}

\section{Introduction}
{Well log data are indispensable in oil and gas exploration, providing pivotal information about the geological formations encountered during drilling \cite{darling2005well,dore2005petroleum,luthi2001geological}. These data are of utmost importance for assessing hydrocarbon potential, petrophysical properties of reservoir rocks, and guiding extraction strategies \cite{alaereeque2014well,ali2022prediction}. However, complex geological conditions can significantly impact the performance of logging equipment \cite{alfakih2023estimating,lai2023typical}, particularly during deep and ultra-deep drilling operations, which frequently encounter high temperatures, high pressures, and highly corrosive geological formations, that can substantially challenge the routine operation of logging equipment \cite{wang2022deep}. Therefore, the integrity of well log data is often compromised by gaps and inaccuracies leading to significant uncertainties in reservoir evaluation and operational decisions \cite{hossain2020missing,liu2021challenges,moore2011uncertainty}.  These challenges are further compounded by a scarcity of available data, underscoring the need for innovative approaches to data handling and analysis in such complex geological settings \cite{alfakih2023reservoir}. 

Current methods for addressing missing or incomplete well log data include traditional statistical imputations and advanced machine learning (ML) techniques \cite{khan2020sice,oluwaseye2022review}. While statistical methods, such as mean imputation and linear regression, are suitable for data with consistent statistical patterns, they struggle with the complex, depth-varying nonlinearity typical of geological data \cite{hallam2022multivariate,lin2020missing,h2022missing}. Recent progress in advanced ML techniques offers improved performance by leveraging the hidden correlations within well log data. Techniques such as decision trees (DTree), random forests (RF), and SVMs have been applied to impute missing data \cite{mcdonald2022impact,feng2021imputation,kurniadi2022local} and generate synthetic datasets with greater accuracy \cite{blanesdeoliveira2021synthetic,yu2021synthetic,akinnikawe2018synthetic}. These methods utilize underlying patterns and relationships within the data to provide more precise imputations than traditional statistical methods.  However, the methods still face significant challenges due to the heterogeneity and inconsistency of geological formations and subsurface conditions (e.g., temperature, pressure, salinity, etc.). This complexity hampers the precise representation of subtle variations within these layers \cite{lai2018review,xu2024numerical}.

Deep learning (DL) methods, a subset of ML, have further advanced the field by introducing models capable of capturing complex temporal and spatial dependencies in well log data. Techniques such as recurrent neural networks (RNNs), long short-term memory networks (LSTMs), gated recurrent units (GRU), bidirectional recurrent neural networks (BRNN), deep neural network- (DNN-), and fully connected neural network (FCNN) have been extensively used for imputation and data generation tasks in time-series data due to their ability to capture temporal dependencies \cite{hallam2022multivariate, zhang2018well, kim2020generation}. For example, bidirectional recurrent imputation for time series (BRITS) leverages bidirectional RNNs to capture dependencies in both forward and backward directions along the time series, significantly enhancing imputation accuracy \cite{cao2018brits}. Similarly, non-autoregressive multiresolution imputation (NAOMI) uses a hierarchical approach to handle imputation at multiple resolutions, accommodating the varied granularities inherent in geological data \cite{liu2019naomi}.

Recent advancements in DL have introduced more sophisticated tools such as transfer learning (TL) that address challenges related to imbalanced multiclass data, sample size, and scalability, further highlighting the versatility and robustness of GAN-based methods in geoscience applications \cite{jamshidi2024synthetic}. Moreover, conditional generative adversarial networks (CGANs) have emerged as pivotal tools in data generation, particularly valued for their capability to incorporate and utilize conditional inputs effectively. This feature enables the precise control of data outputs, making CGANs instrumental across various sectors for tasks such as data imputation and scenario simulation. For instance, in sign language recognition, CGANs enhance accuracy by repairing missing data points based on the probability distribution of such gaps \cite{peng2023missing}. Moreover, combining CGANs with Gaussian mixture models (GMM) clarifies complex nonlinear relationships in seismic data, improving oil saturation predictions \cite{sun2021oil}. 

Recent advancements demonstrate CGANs' utility in integrating with swarm intelligence to optimize well log data imputation, ensuring alignment with geological features \cite{qu2024novel}. Despite these advancements, several challenges remain in well log data imputation and generation, primarily due to the complex, multi-scale nature of geological data. The heterogeneity and inconsistency of geological formations, varying subsurface conditions, and the non-linear relationships within the data make it difficult to achieve accurate imputation and generation. These challenges necessitate the use of more advanced techniques capable of capturing the temporal and spatial dependencies inherent in the data.

To address these challenges, this study introduces an innovative workflow by integrating dual-framework sequence-based GANs. We adapt and leverage time series GAN (TSGAN) \cite{yoon2019timeseries} for generating synthetic well log data and sequence generative adversarial networks (SeqGAN) for the precise imputation of missing values \cite{yu2017seqgan}. The study is enhanced through comparative analyses with advanced deep learning methods like bidirectional recurrent imputation for time series (BRITS) and non-autoregressive multiresolution imputation (NAOMI). This showcases the adaptability and depth of the GAN framework in addressing the complex challenges of geological data analysis. The primary objectives of this study are to develop an enhanced GAN framework tailored for the intricacies of well log data and to evaluate its performance against established models such as BRITS and NAOMI for missing data imputation. The methodology integrates TSGAN for generating synthetic well log data and SeqGAN for precise imputation of missing values. The main contributions of this study are summarized as follows:

\begin{enumerate}
    \item Propose a method based on integrating TSGAN and SeqGAN for well log data analysis, capturing spatial characteristics and correlations to ensure consistency with underlying geological formations.
    \item Conduct an in-depth empirical analysis and evaluation of the proposed methods using different scenarios and evaluation metrics, demonstrating that our approach achieves state-of-the-art results.
\end{enumerate}

\section{Research Method}
The high-level architecture of the proposed framework for well log data analysis begins with data collection, where raw well log data are gathered. This data then undergoes preprocessing, including cleaning, normalizing, handling missing values, and ensuring data integrity as illustrated in Figure 1. The preprocessed data is subsequently used for model training, which is divided into two main tasks: synthetic data creation using TSGAN and data imputation using SeqGAN. The generated synthetic data is validated to ensure it accurately represents the original data's characteristics. The entire process includes a feedback loop from model training back to data preprocessing and synthetic data validation, indicating an iterative refinement approach. The evaluation phase assesses the performance of these models using various performance metrics. Finally, the framework's effectiveness is compared against baseline models to highlight the improvements and advantages of TSGAN and SeqGAN in generating synthetic data and imputing missing well log data, respectively.

\begin{figure}[htbp] % picture
    \centering
    \includegraphics[width=1\textwidth]{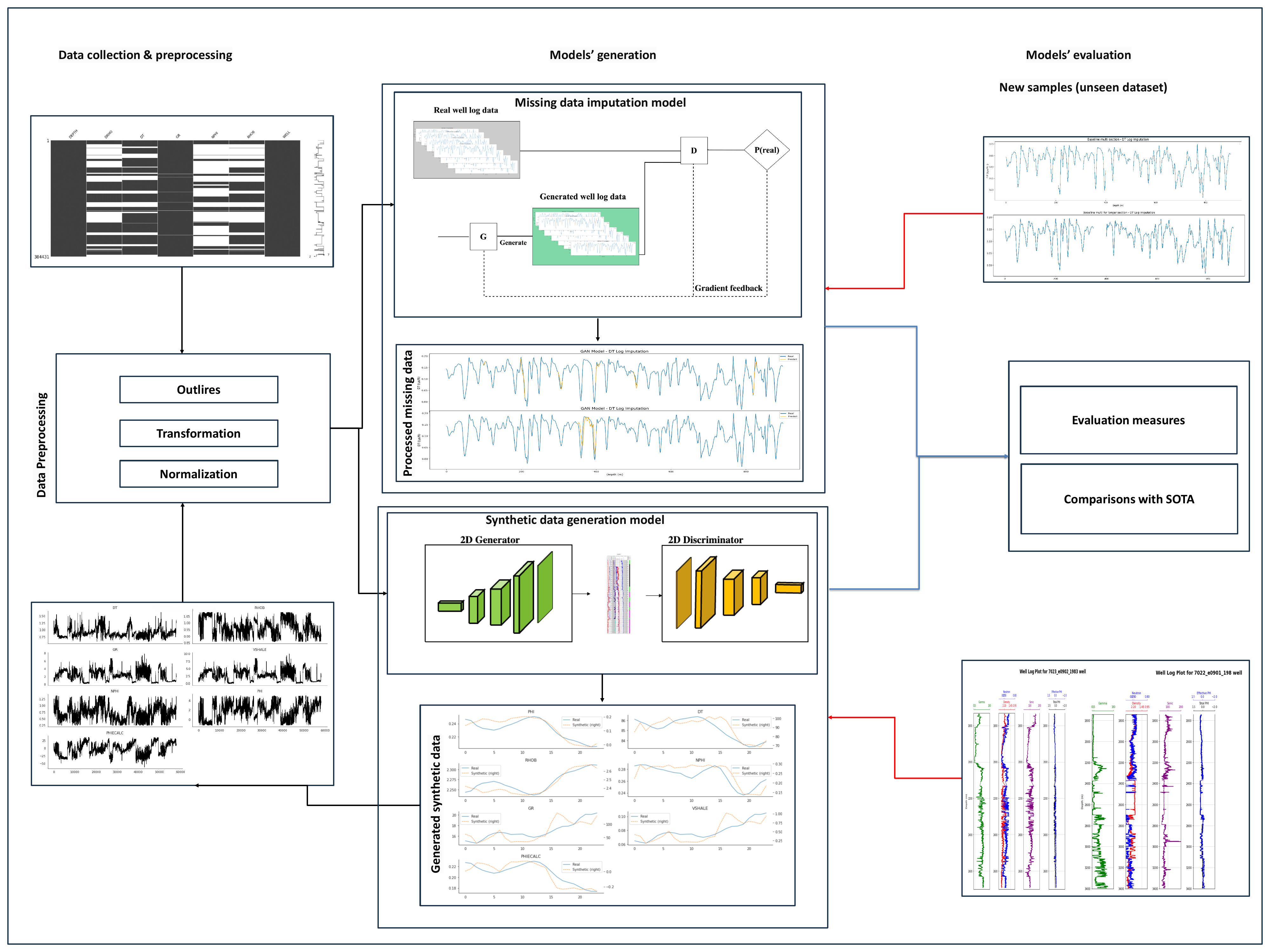}  % Adjust the width as a fraction of the text width
    \caption{High-level architecture of the proposed framework.}

\end{figure}

\subsection{Dataset}
The dataset was sourced from the publicly accessible Netherlands subsurface database (\url{https://www.nlog.nl/en}), focusing on the North Sea Dutch region—a pivotal oil and gas reserve with Jurassic, Cretaceous, and Tertiary formations. The dataset comprised LAS files of well logs, specifically including gamma ray (GR), sonic (DT), neutron porosity (NPHI), deep induction log (ILD), and bulk density (RHOB), selected based on availability.

\subsection{Synthetic well log generation using TSGAN}
\subsubsection{Proposed workflow}
This phase enhances data richness and variability, which is crucial for robust geological analysis. The process, organized into six distinct phases as illustrated in Figure 2, begins with collecting original well logs from the North Sea Dutch area. The data undergoes feature engineering to extract or enhance key features for modeling. Data preprocessing is performed to clean and normalize the data, followed by segmentation to suit model inputs. Correlation analysis between different data attributes helps in fine-tuning feature selection and architectural decisions, ensuring the data is optimally prepared for modeling.

The TSGAN architecture includes an autoencoder that compresses data into a dense latent space capturing requisite characteristics. The GAN setup comprises a discriminator and a generator. The discriminator evaluates the authenticity of data samples while the generator creates synthetic samples indistinguishable from real ones. The model processes both static and temporal features, enhancing its capability to handle complex depth series data efficiently. During the training process, the autoencoder is refined to minimize reconstruction errors, improving latent space representations. Supervised and joint training sessions help align synthetic data closely with real data statistics and spatial dynamics, which is crucial for generating high-quality synthetic sequences. 

Embedding functions transform the data into a suitable format for training, focusing on key feature extraction and dimensionality reduction. A final training loop through adversarial training fine-tunes both the generator and discriminator, enhancing their overall performance. Synthetic time-series data are generated to mimic the statistical properties of the original dataset. This synthetic data is then rescaled and plotted to visually compare the outputs against the real data, ensuring quality and consistency. Finally, the data is visualized using principal component analysis (PCA) and t-distributed stochastic neighbor embedding (t-SNE). These techniques simplify and visualize the complexity of the data, providing a clear graphical representation of how synthetic data compares to real data. This step ensures the synthetic data's utility and reliability for geophysical analysis.

\begin{figure}[htbp] % picture
    \centering
    \includegraphics[width=1\textwidth]{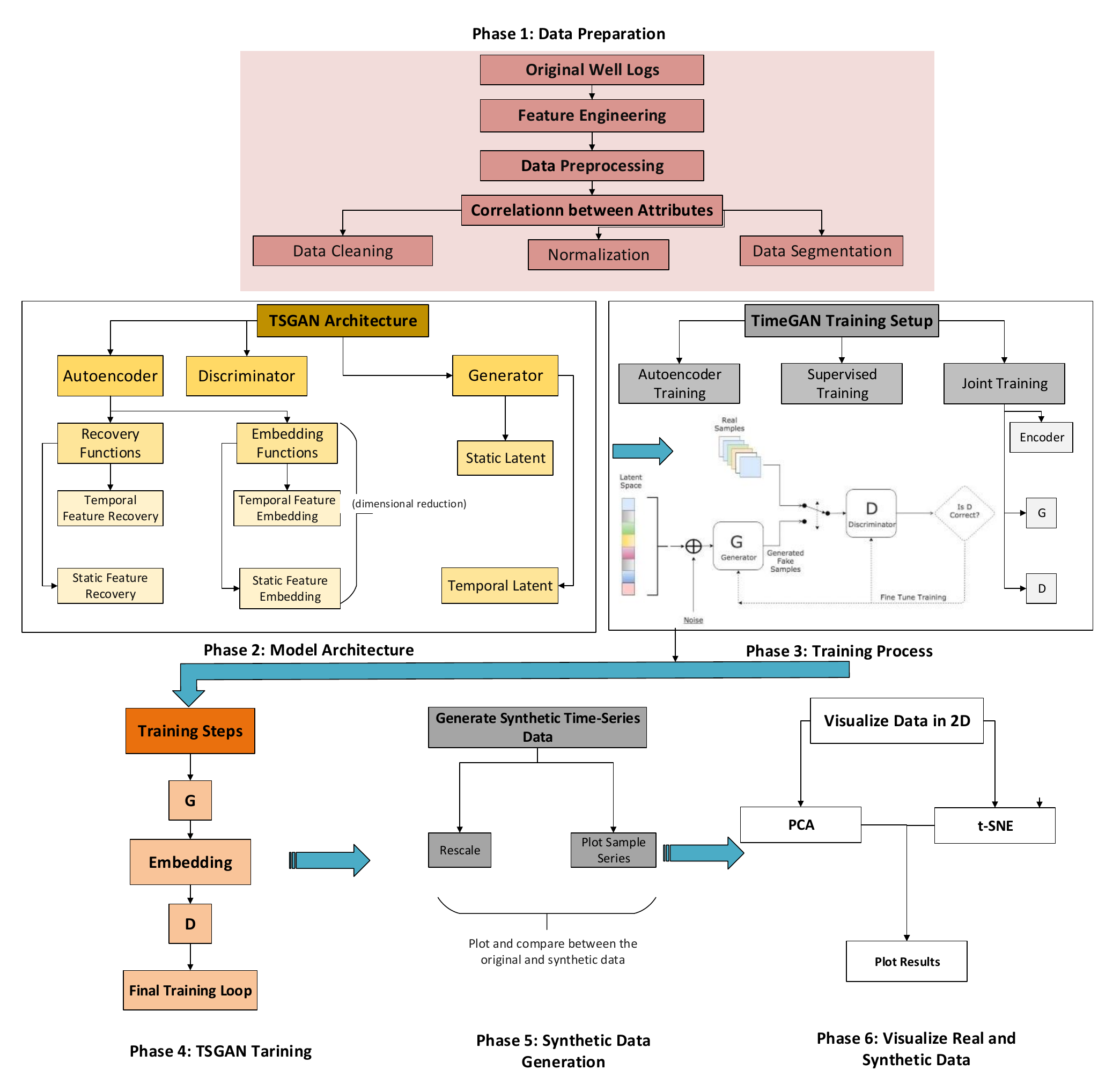}  % Adjust the width as a fraction of the text width
    \caption{TSGAN workflow for synthetic data generation.}
    
\end{figure}

 \subsubsection{Time Series generative adversarial networks (TSGAN)}

TSGAN offers a robust framework for generating realistic time-series data by capturing the inherent spatial dynamics of sequential data. Traditional GANs face challenges in maintaining the sequential dependencies necessary for time-series data. Although originally developed for time series, we have adapted TSGAN for spatial sequences in this study. TSGAN addresses these challenges by integrating both unsupervised and supervised learning paradigms within a generative adversarial architecture, thereby preserving the spatial correlations fundamental for realistic synthetic data generation \cite{chen2018modelfree, huang2023deep, li2019madgan, yoon2019timeseries}.

In this study, TSGAN is utilized for generating synthetic depth-series data. Figure 3 illustrates the fundamental components of a TSGAN architecture, where the generator (G) attempts to create data that mimics real sequences and the discriminator (D) evaluates the authenticity of this data. This interaction forms the core of GAN operations, setting a foundation for more complex adaptations.
\begin{figure}[htbp] % picture
    \centering
    \includegraphics[width=1\textwidth]{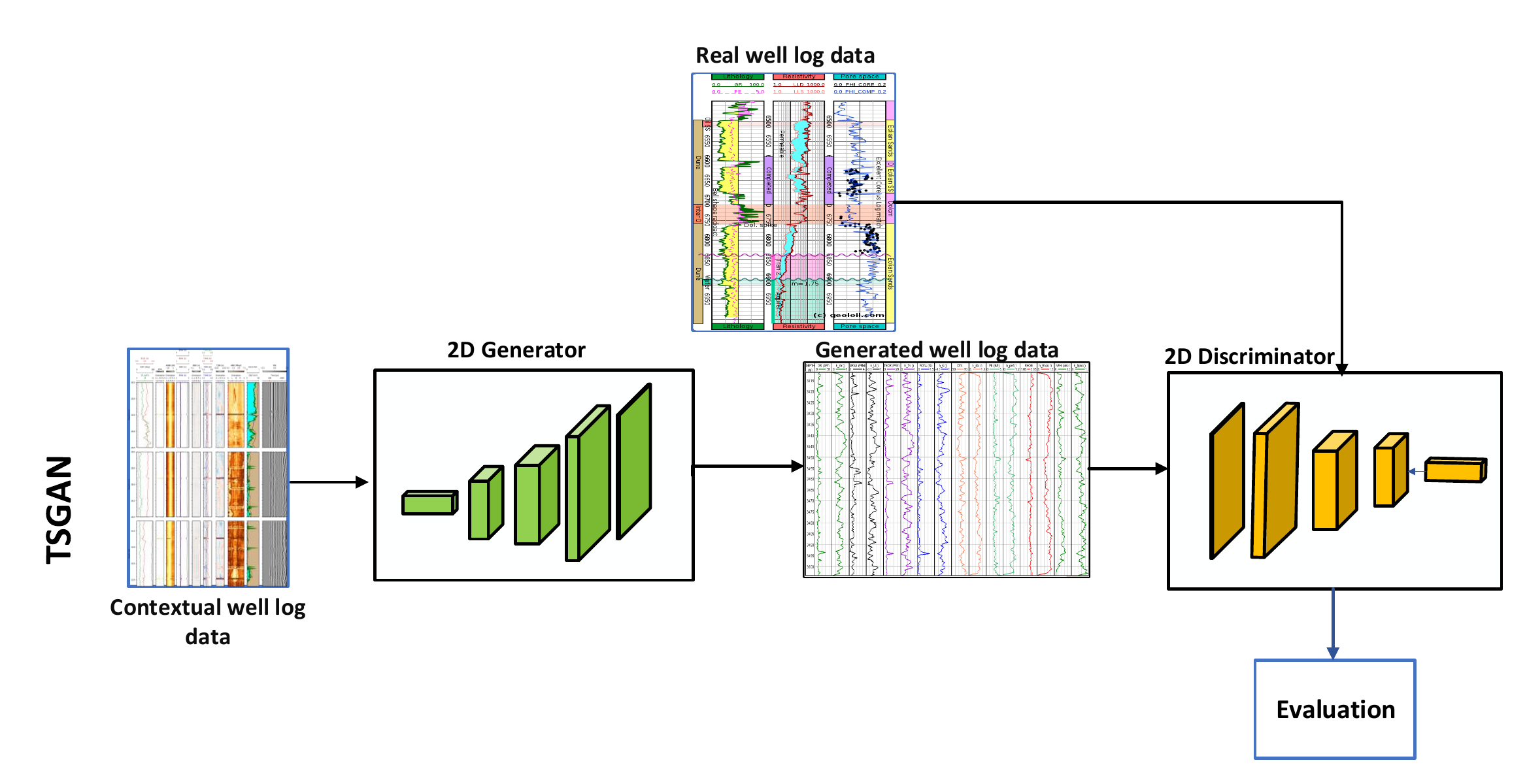}  % Adjust the width as a fraction of the text width
    \caption{TSGAN architecture.}
    
\end{figure}

Using TSGAN is motivated by its proven ability to handle complex data distributions effectively \cite{vera2023protect, sharma2024gan}. TSGAN is particularly well-suited for managing temporal/spatial sequences characteristic of well log data, ensuring that the synthetic data maintains realistic time-series properties \cite{yoon2019timeseries}. The architecture includes a two-stage process: the first stage involves generating the primary features of the data, while the second stage refines these features to improve the accuracy and realism of the synthetic data. This dual-stage approach helps in capturing intricate patterns and dependencies in the data, making TSGAN highly effective for this application.

TSGAN is constructed from several key components:
\begin{itemize}
    \item \textbf{Generator (G)}: The generator creates time-series data trying to mimic the real data distribution.
    \item \textbf{Discriminator (D)}: The discriminator distinguishes between the outputs of the generator and the actual data.
    \item \textbf{Embedding Network}: This transforms high-dimensional time-series data into a lower-dimensional, manageable latent space. This transformation aids in capturing the crucial features of data while reducing noise.
    \item \textbf{Recovery Network}: This maps the latent representations back to the original data space, ensuring that the generated sequences can be translated back into interpretable time-series data.
\end{itemize}

\subsubsection{Mathematical formulation of TSGAN}
The training of TSGAN is structured around the optimization of both adversarial and supervised losses:

\begin{itemize}
    \item \textbf{Adversarial Loss ($L_U$)}: Ensures that the discriminator cannot easily distinguish between real and generated data, pushing the generator towards producing more realistic sequences.
    \item \textbf{Supervised Loss ($L_S$)}: Involves comparing the generated sequences against real data sequences to ensure accurate replication of spatial dynamics.
    \item \textbf{Moment Loss ($L_M$)}: Focuses on matching the statistical moments of the real and synthetic data. In this study, we consider the 1st (mean) and 2nd (variance) moments.
\end{itemize}

The total loss function is given by $L = L_U + \lambda L_S$, where $\lambda$ is a weighting factor that balances the importance of the unsupervised and supervised components.

The following expressions describe the various mathematical operations within TSGAN:

\begin{itemize}
    \item \textbf{Embedding function:} Maps input sequences into a latent space.
    \[
    e(s, x_{1:T}) \rightarrow (h_s, x_{h_{1:T}}) \tag{1}
    \]
    Equation (1) describes the transformation of the real sequence $s, x_{1:T}$ into latent codes $(h_s, h_{1:T})$.
    
    \item \textbf{Generator function:} Generates synthetic sequences from random noise conditioned on latent codes.
    \[
    g(z, z_{1:T}) \rightarrow (h_s', h_{1:T}') \tag{2}
    \]
    Equation (2) represents the generation of synthetic sequences $(h_s', h_{1:T}')$ from latent variables $g(z, z_{1:T})$.
    
    \item \textbf{Discriminator function:} Differentiates between real and synthetic sequences.
    \[
    d(h_s, h_{1:T}) \rightarrow y \tag{3}
    \]
    Equation (3) illustrates how the discriminator outputs a classification $y$ based on whether the sequence is deemed to be real or synthetic.
    
    \item \textbf{Recovery function:} Recovers synthetic sequences back to their original data space.
    \[
    r(h_s, h_{1:T}) \rightarrow (s_s', h_{x_{1:T}}') \tag{4}
    \]
    Equation (4) details the recovery of the data space sequences $(s_s', h_{x_{1:T}}')$ from the synthetic latent codes. In the above expressions, $T$ denotes the total length of the sequence vector.
\end{itemize}
\subsubsection{Broader applications of TSGAN}
 TSGAN has been widely applied to generate synthetic data across various applications, showcasing its versatility and robustness in handling complex datasets. In environmental modeling, TSGAN was used in a Bayesian GAN approach to predict combined sewer flow, demonstrating its potential in urban planning and environmental engineering \cite{bakhshipour2023bayesian}. In the healthcare sector, it has enhanced clinical diagnostics and patient monitoring by augmenting sensor-based health data and improving medical model robustness \cite{yang2023tsgan}. Additionally, TSGAN's application in real-time data analysis has been instrumental in increasing accuracy and reliability in time-sensitive systems by integrating synthetic with real-time data \cite{juneja2023synthetic}. In the field of knowledge discovery, it has been used to enhance data privacy and improve the quality of synthetic data for ML models, as demonstrated in a recent study, which highlighted its effectiveness in generating high-fidelity time-series data that maintain statistical similarities to original datasets \cite{klopries2024itfgan}. Similarly, a recent preprint details TSGAN's use for augmenting training datasets in healthcare, particularly for ECG data analysis, further underscoring its capability to support better patient outcomes through realistic, diverse biomedical time-series data generation \cite{vera2023protect}.

\subsection{Well log imputation using SeqGAN BRITS, and NAOMI}
\subsubsection{Proposed workflow}
The workflow aims to restore data integrity by accurately filling in missing data segments, ensuring datasets are comprehensive and reliable for subsequent analysis. Figure 4 outlines the sequential steps involved in data imputation using SeqGAN, BRITS, and NAOMI, demonstrating a structured approach to address the challenges of missing data in well logs. The process begins in Phase 1, where well logs are prepared by segmenting the original data into training and testing datasets, ensuring that both datasets adequately represent the complexities of real-world geological formations. This phase sets the foundation for effective model training by establishing a diverse data environment. 

In Phase 2, each model undergoes rigorous training tailored to its unique capabilities. SeqGAN leverages adversarial training to refine the generation and discrimination of sequential data, enhancing its ability to replicate and restore spatial sequences. BRITS utilize bidirectional recurrent neural networks to exploit spatial dependencies effectively, ensuring the continuity and coherence of time-series data. Simultaneously, NAOMI applies a hierarchical approach to tackle imputation at multiple resolutions, accommodating the varied granularities inherent in geological data.

Phase 3 transitions to testing and validation, where models are subjected to real-world conditions simulated by the application of mask matrices. This phase is imperative as it involves the application of the models to the testing dataset to fill in missing values and verify their imputation accuracy. The use of mask matrices helps simulate various scenarios of missing data, providing a robust test environment to evaluate each model's effectiveness.

Finally, Phase 4 focuses on the refinement and evaluation of the models. It involves fine-tuning the models based on the feedback received from the testing phase. This step is crucial for optimizing the models to enhance their accuracy and reliability in predicting and restoring missing well log data. The continuous iteration and refinement help in achieving the highest standards of data quality, essential for precise geological analysis.
\begin{figure}[htbp] % picture
    \centering
    \includegraphics[width=1\textwidth]{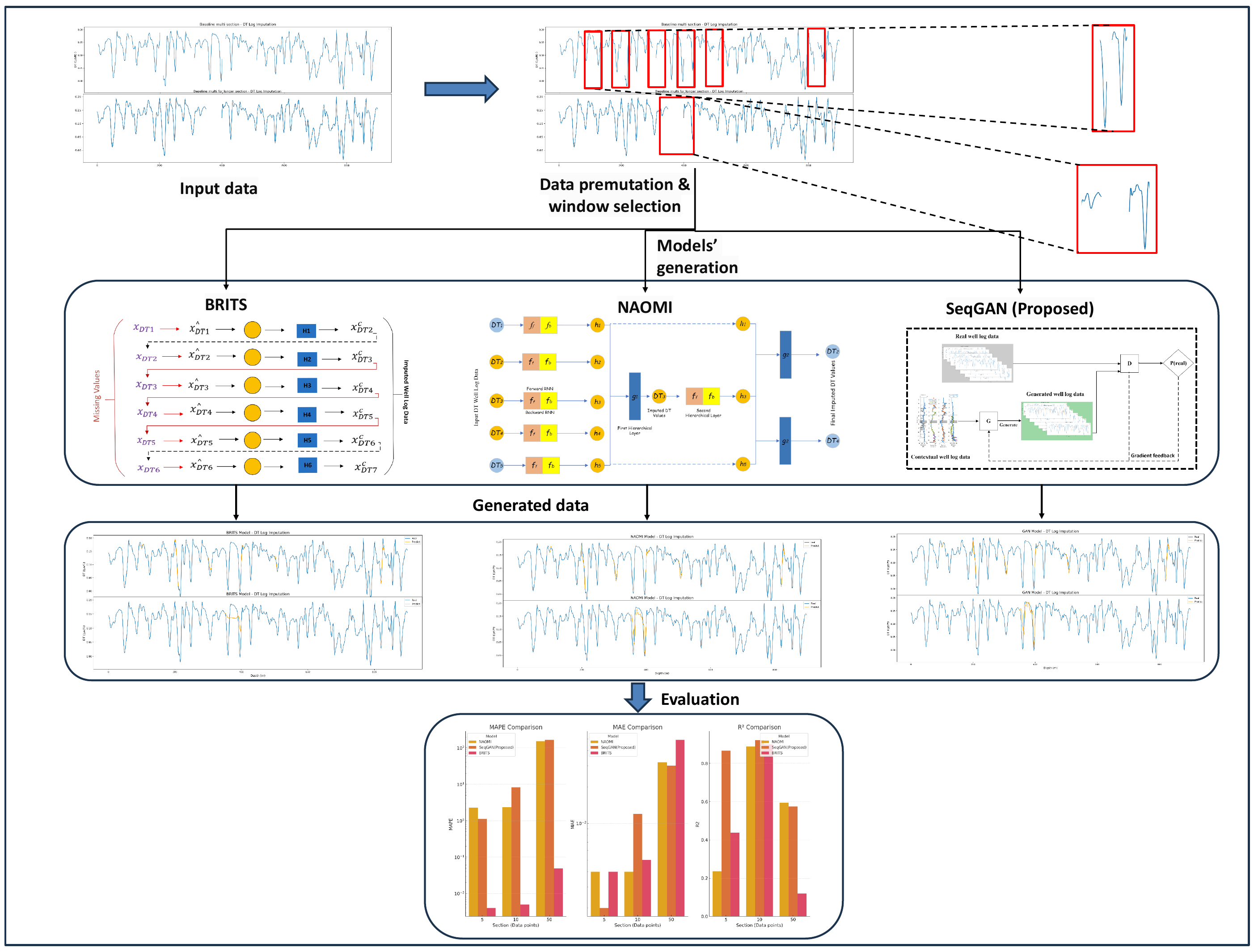}  % Adjust the width as a fraction of the text width
    \caption{Workflow for log imputation.}
    \label{1}
\end{figure}

\subsubsection{Sequential generative adversarial networks (SeqGAN)}
In addressing the critical issue of missing data in well log sequences, SeqGAN emerges as a specialized solution that surpasses traditional GANs by focusing specifically on the sequential nature of the data. SeqGAN adapts the adversarial training framework to effectively impute missing sequences, ensuring that the continuity and spatial dependencies integral to geological sequences are maintained. Figure 5 (a) illustrates the SeqGAN framework, showing how it intelligently fills in missing sequence data. The generator in SeqGAN is trained to predict missing segments based on both preceding and subsequent available data, while the discriminator assesses the coherence and authenticity of the generated sequences against real data. This dynamic interaction ensures that the imputed data are not only plausible but also consistent with the spatial patterns of existing sequences.

SeqGAN is particularly valuable for well log data analysis, where maintaining the integrity of depth-series data is crucial. The ability to accurately reconstruct missing parts of a sequence greatly enhances data reliability and usability for subsurface geological analysis. SeqGAN's approach provides a robust method for dealing with the often irregular and gap-ridden data obtained from field measurements, ensuring that subsequent analyses and decisions are based on comprehensive and accurate data sets \cite{yu2017seqgan}. Unlike traditional imputation methods that might treat data points independently or apply simple interpolation, SeqGAN utilizes the sequential context of data, which is crucial for depth-series like well logs. This contextual awareness allows SeqGAN to produce more accurate and realistic imputations, particularly in complex scenarios where the data dependencies are significant.

In the landscape of missing data imputation within well logs, the BRITS and NAOMI models stand out for their sophisticated approaches to handling temporal/spatial and multi-resolution data respectively.

BRITS leverages the power of bidirectional recurrent neural networks (RNNs) to capture dependencies in both forward and backward directions along the sequence. This bidirectional approach is particularly effective in contexts where deeper data points influence shallower imputation, common in sequential data like well logs. Figure 5 (b) illustrates this bidirectional flow, emphasizing how BRITS processes spatial dependencies to enhance the accuracy and integrity of the imputed data.

NAOMI, on the other hand, adopts a hierarchical approach to address imputation at multiple resolutions, which is crucial for well log data characterized by varying granularities and depth scales. This model progressively refines its imputations, starting from coarse resolutions and moving towards finer details, thereby improving the overall data quality for complex geological formations. Figure 5 (c) provides a schematic of this process, showing the layering and multi-resolution handling that distinguishes NAOMI.
\begin{figure}[htbp]
    \centering
    \begin{minipage}{1\textwidth}
        \centering
        \includegraphics[width=\textwidth]{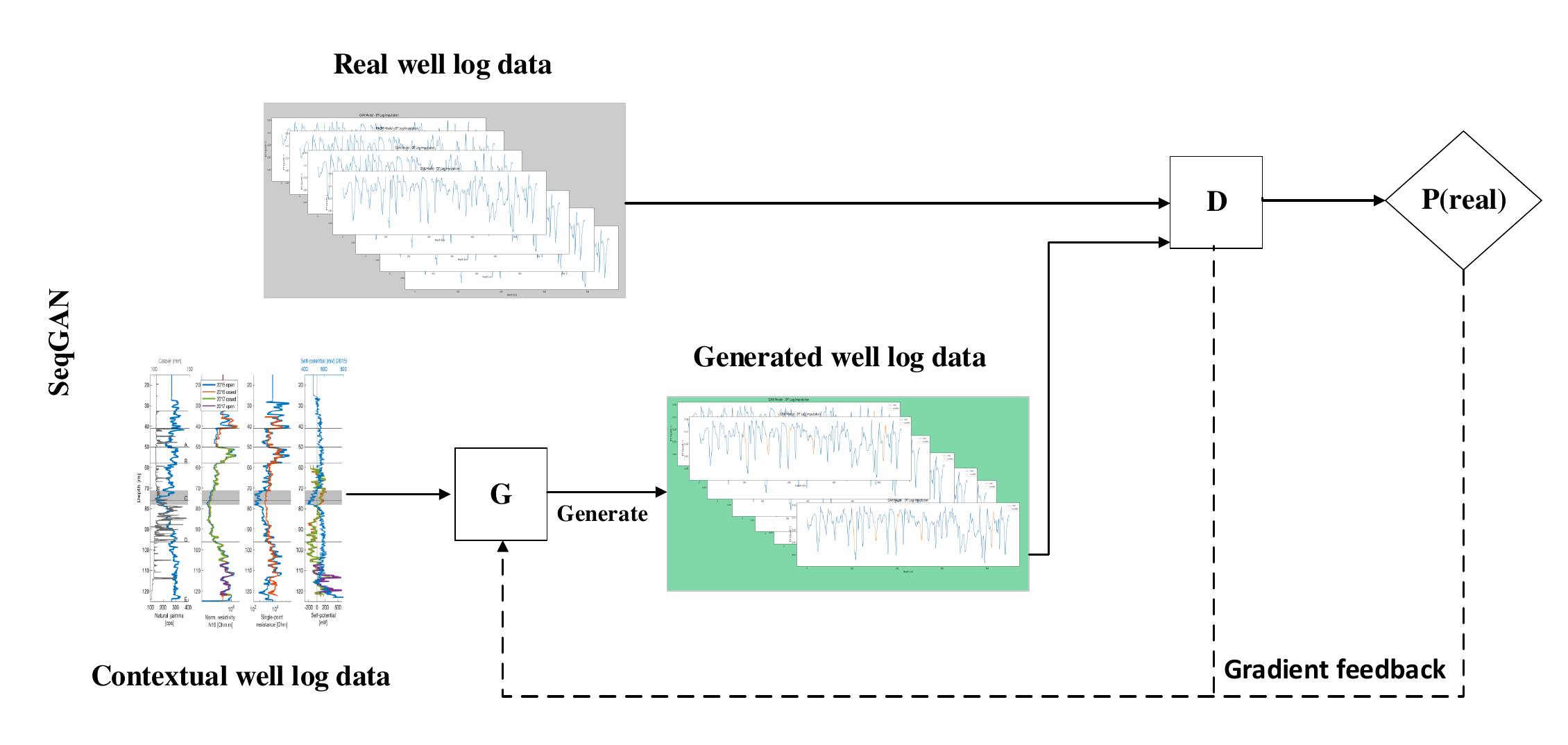}  % Image for (a)
        \subcaption{SeqGAN Architecture}
        \label{1a}
    \end{minipage}\hfill
    \begin{minipage}{0.8\textwidth}
        \centering
        \includegraphics[width=\textwidth]{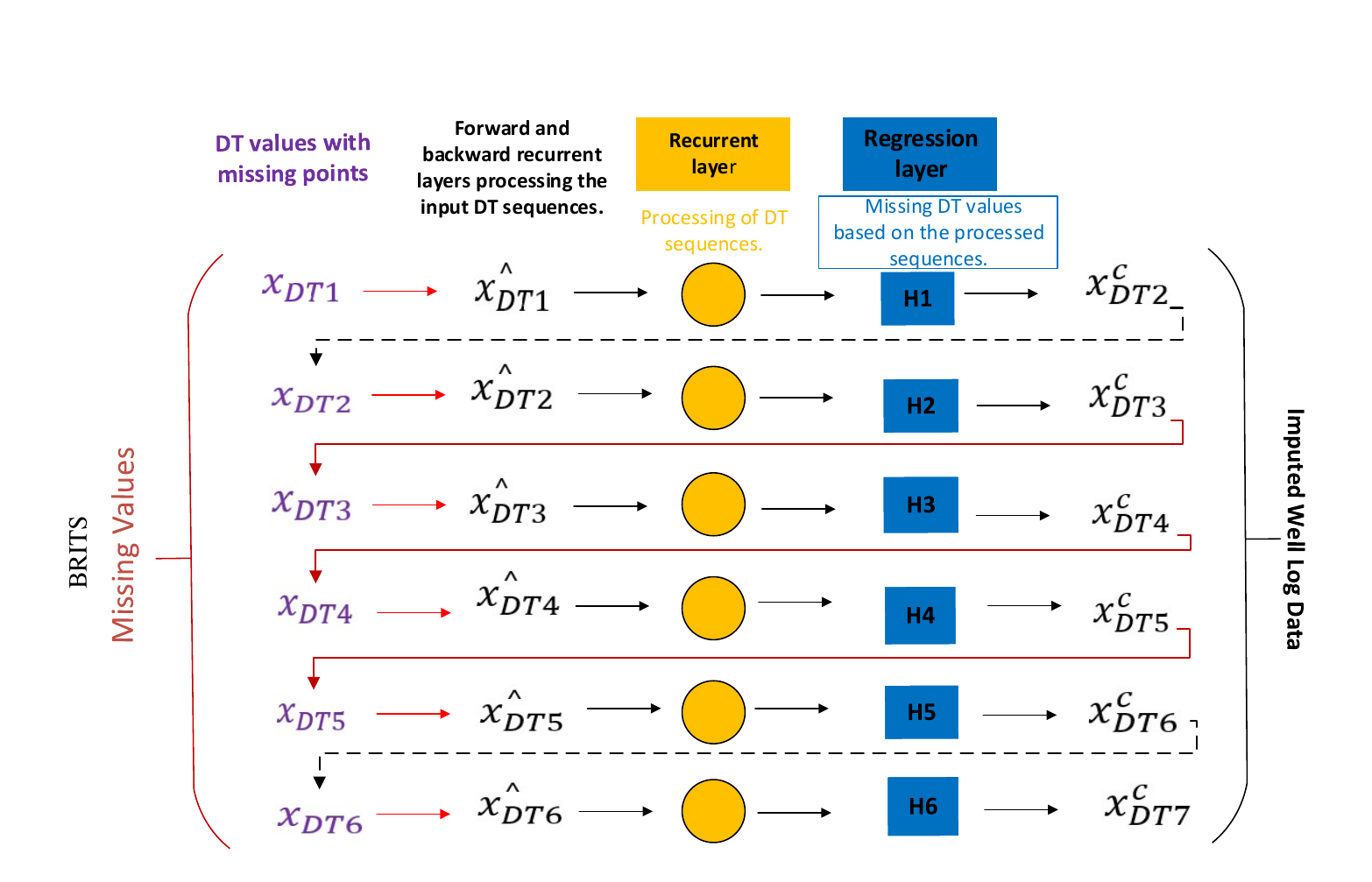}  % Image for (b)
        \subcaption{BRITS Architecture}
        \label{2b}
    \end{minipage}\hfill
    \begin{minipage}{1\textwidth}
        \centering
        \includegraphics[width=\textwidth]{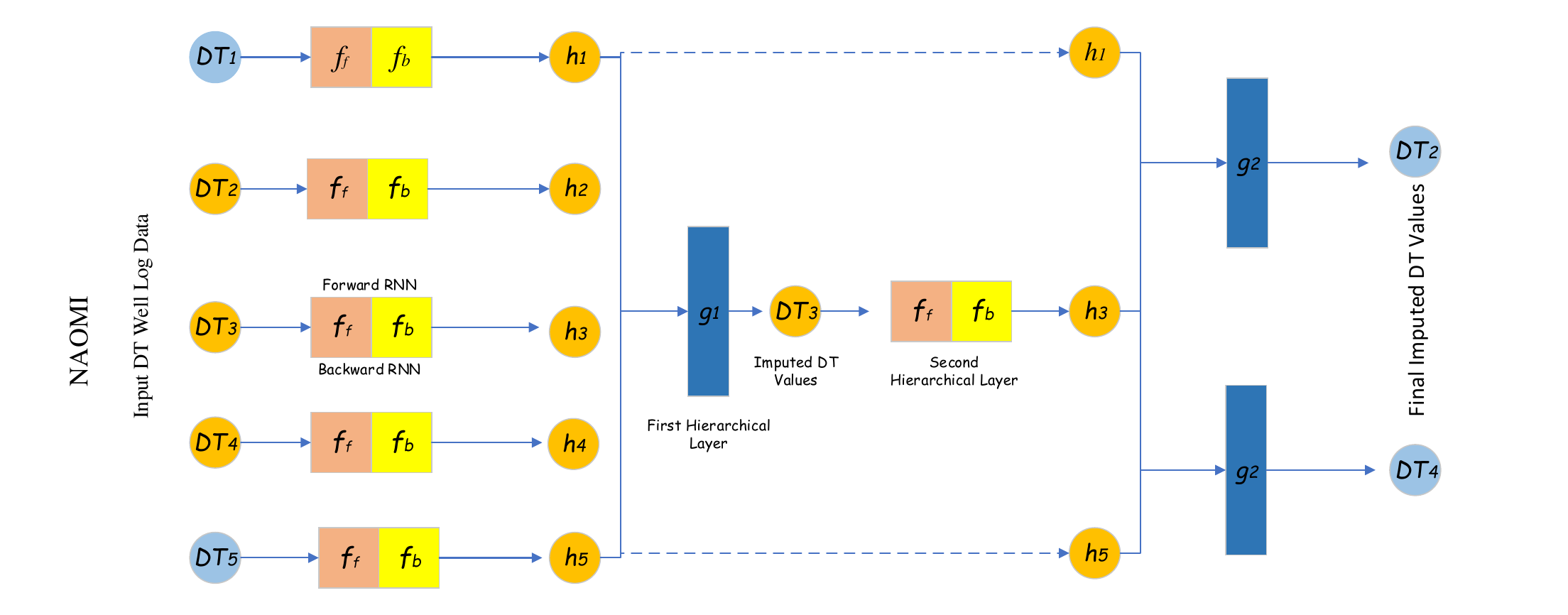}  % Image for (c)
        \subcaption{NAOMI Architecture}
        \label{3c}
    \end{minipage}
    \caption{Workflow for log imputation: (a) SeqGAN, (b) BRITS, and (c) NAOMI architectures.}
    \label{4}
\end{figure}

\section{Experimental work}
\subsection{Dataset preprocessing}
The LAS files were converted to CSV format using Python to simplify data manipulation and visualization. Initial data exploration was conducted using D-Tale, a Python library for visual data analysis, which is an open-source tool designed for exploring and visualizing datasets (\url{https://github.com/andymcdgeo/Petrophysics-Python-Series}). This tool enabled effective identification and visualization of missing data points across the dataset, facilitating a better understanding of the data's structure and quality.

Outlier detection and removal were performed using multiple methods: isolation forest (IF), one-class support vector machine (SVM), and local outlier factor (LOF). The IF method proved to be the most effective, isolating approximately 10\% of data points as outliers per well. We manually tested the IF method to detect outliers in the 5-20\% range and determined that removing approximately 10\% of the data points as outliers was the best choice, as it provided a logical balance between data retention and accuracy. This refined outlier detection process was applied across all well logs, including gamma ray (GR), sonic (DT), neutron porosity (NPHI), bulk density (RHOB), and deep induction log (ILD). Outliers are marked as orange points, while inliers are marked as blue points. This approach was crucial for maintaining the dataset's quality and enhancing the accuracy and reliability of subsequent modeling efforts.

Appendix Figure 11A shows the three methods used for outlier detection: SVM, IF, and LOF. Appendix Figure 12A demonstrates the application of the IF method to all well logs, illustrating the identification and removal of approximately 10\% of the data points as outliers. Appendix Figure 13A provides a visual comparison of well log data before and after cleaning using the IF method, highlighting the enhancement of data accuracy and model reliability for reservoir property predictions. This comprehensive approach to outlier detection and removal was crucial for maintaining the dataset's quality and enhancing the accuracy and reliability of subsequent modeling efforts.

To underscore the initial challenges encountered with the dataset, Figure 6 illustrates the distribution of missing data across various well log parameters such as GR, DT, NPHI, ILD, and RHOB. Each column in the figure represents a specific type of well log, with white spaces indicating missing data points across different depths. The separate column on the extreme right of Figure 6, with the x-axis labeled 2-7, represents the well identifiers. Each unique value on this axis corresponds to a specific well from which the data was collected. This column helps distinguish the data points based on their source wells, providing additional context for the distribution and gaps observed in the well log parameters.

This visualization was generated using D-Tale, an advanced Python library for data analysis, which facilitated the detailed exploration and identification of missing entries. Identifying these missing points was instrumental in tailoring the imputation strategies such as TSGANs, BRITS, and NAOMI, chosen not only for their ability to impute missing data but also to preserve the integrity of geological data crucial for accurate subsurface analysis.

\begin{figure}[htbp] % picture
    \centering
    \includegraphics[width=1\textwidth]{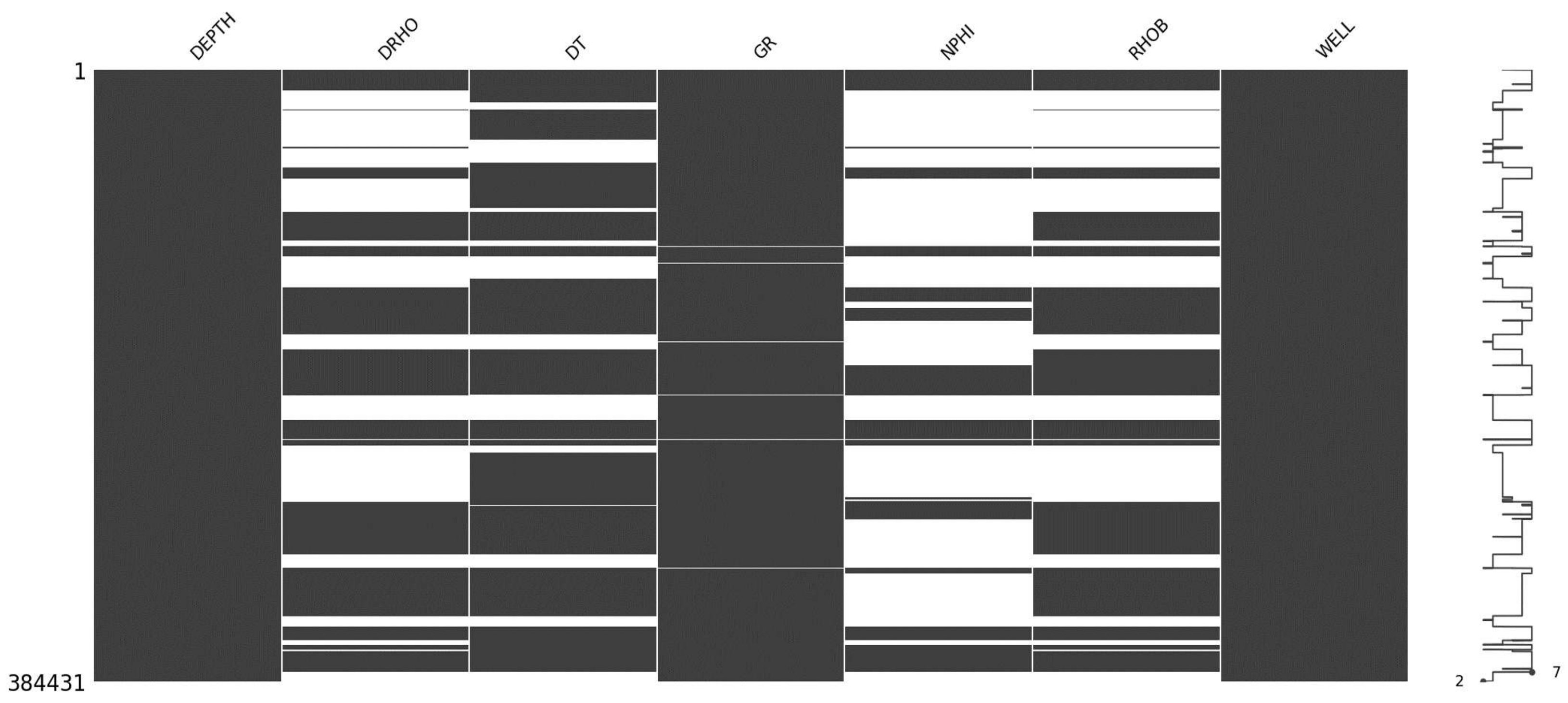}  % Adjust the width as a fraction of the text width
    \caption{Distribution of missing values in well log data across various parameters (GR, DT, ILD, NPHI, and RHOB). The separate column on the extreme right indicates the well identifiers (2-7), showing the source of the data points and facilitating the understanding of data distribution across different wells.}
    \label{1}
\end{figure}

For the training and test dataset selection, the well log data was split into training and validation sets, with the validation set comprising 20\% of the data. Specific depth ranges were intentionally corrupted to simulate realistic missing data scenarios, creating test datasets with gaps at intervals such as 100-105m, 200-210m, and so on. The preprocessed data was normalized using the MinMaxScaler, which scales values between 0 and 1, ensuring compatibility with machine learning algorithms. 

Sliding windows of fixed sizes were employed to create batches that captured local continuity and patterns within the logs. The sequences were standardized using the StandardScaler, which removes the mean and scales the data to unit variance. The NAOMI and BRITS models were trained on these sequences, learning to predict the missing values by analyzing the surrounding log values. 

The models' performance was evaluated using metrics such as mean absolute percentage error (MAPE), mean absolute error (MAE), and R-squared (R²), comparing the imputed values against the actual values in the test datasets.

\subsection{Real log data visualization}
To illustrate the raw data with missing points, we selected examples from the NPHI and RHOB logs, as these parameters exhibited the highest proportion of missing data, as shown in Figure 7. The visualization of these logs before imputation helps underscore the extent of the missing data problem and the necessity for effective imputation techniques.
\begin{figure}[htbp]
    \centering
    \begin{minipage}{1\textwidth}
        \centering
        \includegraphics[width=\textwidth]{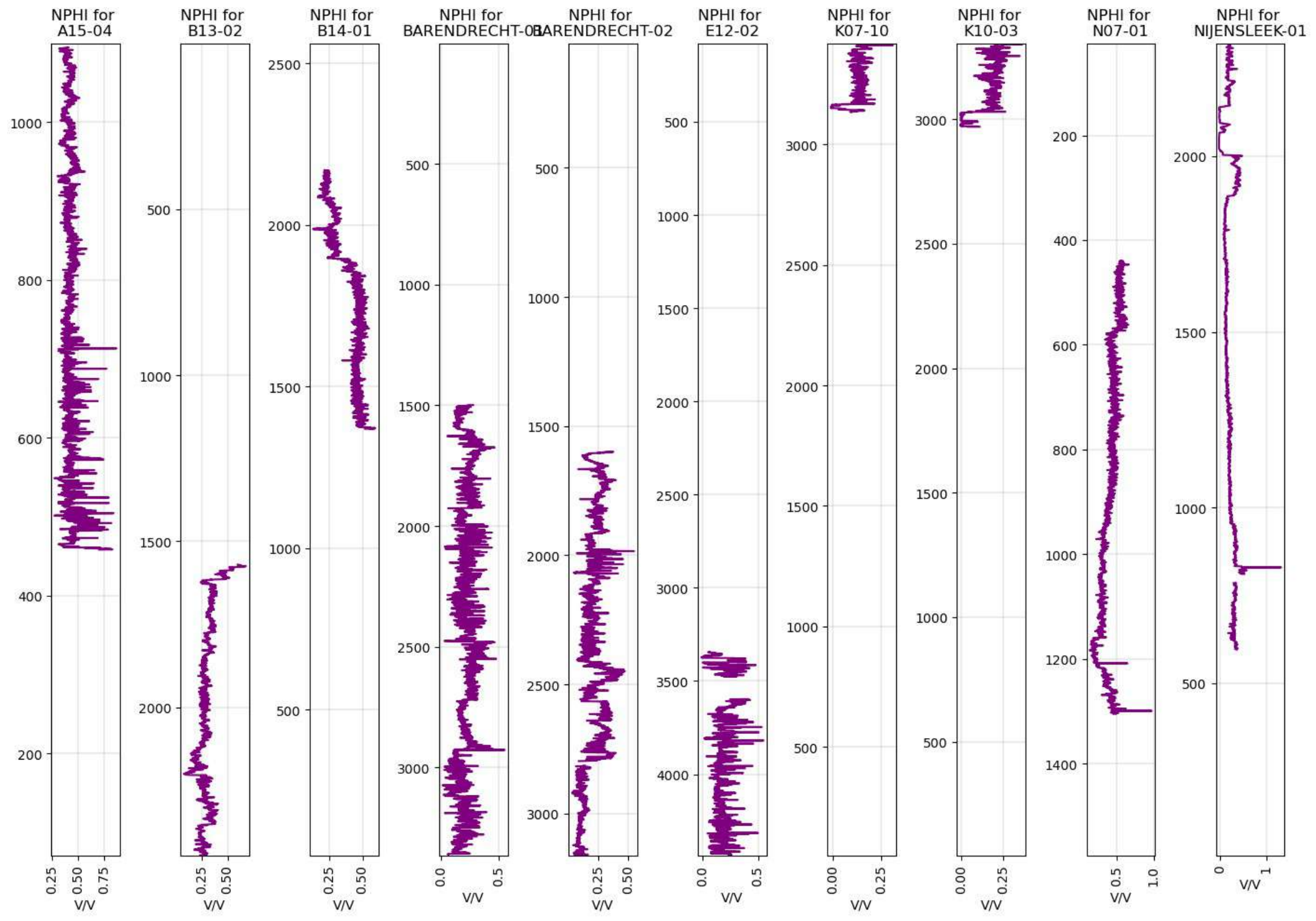}  % Image for (a)
        \subcaption{Real NPHI logs for multiple wells, illustrating the raw data with missing points.}
        \label{1a}
    \end{minipage}\hfill
    \begin{minipage}{1\textwidth}
        \centering
        \includegraphics[width=\textwidth]{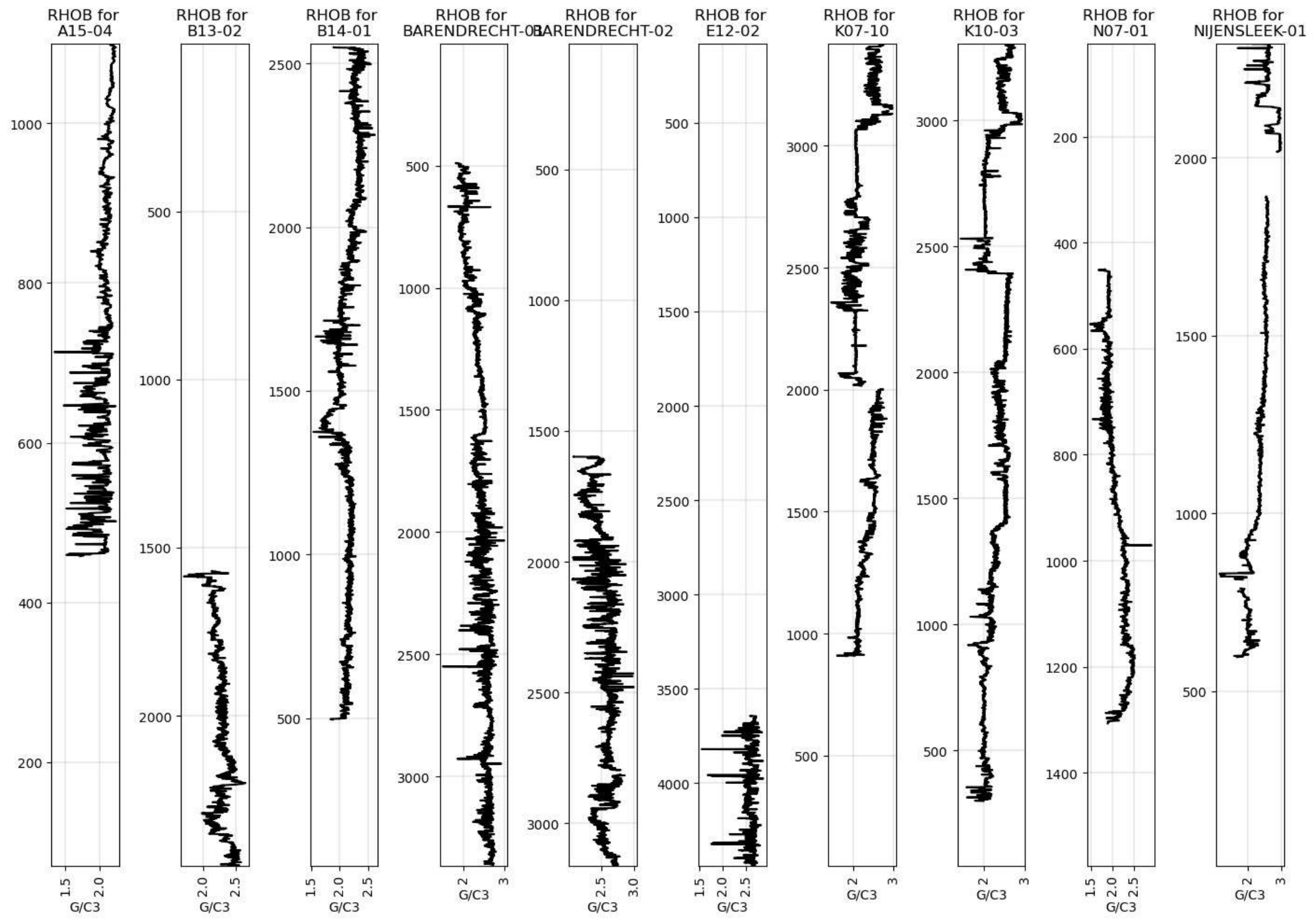}  % Image for (b)
        \subcaption{Real RHOB logs for multiple wells, illustrating the raw data with missing points.}
        \label{2b}
    \end{minipage}\hfill
    \caption{Real log data for multiple wells, illustrating the raw data with missing points for NPHI and RHOB parameters. These visualizations highlight the extent of missing data and the necessity for effective imputation techniques.}
    \label{4}
\end{figure}

\subsection{Experiment settings}
\subsubsection{Synthetic generation model configuration and parameters}
Table 1 summarizes the architecture and hyperparameters of the TSGAN model, which is crucial for synthesizing time-series data. The model components include the embedder, recovery, generator, supervisor, and discriminator, all leveraging gated recurrent unit (GRU)-based architectures for capturing spatial dependencies.

Hyperparameters, like hidden layer size and learning rate, are carefully chosen to balance model complexity and training stability. A consistent hidden size of 24 ensures uniformity in latent space representation, while optimization algorithms like the Adam optimizer and loss functions such as binary cross-entropy (BCE) and mean squared error (MSE) adhere to GAN training best practices. The TSGAN model employs three primary loss functions, each contributing to different components of the model:

\begin{itemize}
    \item \textbf{Adversarial Loss:} Used for training both the generator and discriminator. The BCE loss ensures that the discriminator correctly distinguishes between real and synthetic data and that the generator produces data indistinguishable from real data \cite{goodfellow2014generative}.
    \item \textbf{Supervised Loss:} Applied to the supervisor and the generator. The MSE loss ensures that the generated data maintains the temporal dependencies of the real data, helping the supervisor learn sequence-to-sequence relationships effectively \cite{yoon2019timeseries}.
    \item \textbf{Reconstruction Loss:} Used in the autoencoder (embedding and recovery functions) to minimize the difference between the input real data and the recovered data. The MSE loss ensures accurate reconstruction of the original data from its latent representation \cite{yoon2019timeseries}.
\end{itemize}

The total loss for training the TSGAN model is a weighted sum of these losses:

\[
\text{Generator Loss} = \text{Adversarial Loss} + \text{Supervised Loss} + 100 \times \sqrt{\text{Supervised Loss}} \tag{5}
\]

The \textbf{Discriminator Loss} primarily consists of the adversarial loss, ensuring that the discriminator can effectively differentiate between real and synthetic data. The discriminator loss is also calculated using the BCE loss \cite{goodfellow2014generative}. 

Parameters like batch size and sequence length determine input data granularity and length, which are essential for capturing long-term dependencies and generating coherent outputs. Additional parameters like the number of training steps, perplexity (for t-SNE), and the number of components (for PCA) offer control over training and evaluation, facilitating comprehensive analysis.

\begin{table}[h]
  \caption{TSGAN model configuration and hyperparameters.}
  \centering
  \resizebox{\textwidth}{!}{
  \begin{tabular}{|l|l|l|l|}
    \hline
    \textbf{Hyperparameter} & \textbf{Description} & \textbf{Value} & \textbf{Notes} \\
    \hline
    Model Components & Architecture for all & GRU-based & Captures spatial dependencies \\
    \hline
    Adam Optimizer & Training model parameters & Learning Rate: 0.001 & Optimizes model training \\
    \hline
    BCE & Discriminator loss & Applied in discriminator and adversarial training & Ensures discriminator accuracy \\
    \hline
    MSE & Reconstruction loss & Applied in autoencoder and supervisor & Ensures accurate reconstruction and sequence learning \\
    \hline
    Batch Size & Samples processed per iteration & 128 & Important for training efficiency \\
    \hline
    Sequence Length & Length of input sequences & 24 & Controls captured spatial scope \\
    \hline
    Learning Rate & Rate of parameter updates & 0.001 & May require tuning \\
    \hline
    Number of Training Steps & Total training iterations & 20000 & May require adjustment \\
    \hline
    Perplexity (t-SNE) & Parameter for dimensionality reduction & 30 & Controls t-SNE complexity \\
    \hline
    Number of Components (PCA) & Number of principal components (if applicable) & 24 & Controls captured variance \\
    \hline
  \end{tabular}
  }
  \label{tab:tsgan_config}
\end{table}

\subsubsection{Imputation model configuration and parameters }
After detailing the specific architecture of each generative model employed in this study, Table 2 presents a comprehensive breakdown of the training parameters and configurations used for each. This comparative overview not only underscores the tailored approaches taken to optimize each model but also highlights the adaptability and specificity of the methods in dealing with the unique challenges presented by well log data. Selecting appropriate batch sizes, learning rates, and iteration counts was vital to balancing computational efficiency with the learning capabilities of each model. For instance, the higher iteration count in GAN training reflects its sensitivity to discriminator convergence, a crucial aspect for generating high-fidelity synthetic data.

\begin{table}[h]
  \caption{Comparison of training parameters across the NAOMI, GAN, and BRITS models detailing epochs, batch sizes, learning rates, and other important settings employed in the study. This overview provides insights into the model configurations tailored to enhance performance and accuracy in well log data imputation.}
  \centering
  \resizebox{\textwidth}{!}{
  \begin{tabular}{|l|l|l|l|}
    \hline
    \textbf{Parameter} & \textbf{NAOMI} & \textbf{SeqGAN} & \textbf{BRITS} \\
    \hline
    Epochs/Iterations & 200 epochs & Pretrain: 2000 iterations, Train: 1000 iterations & 200 epochs \\
    \hline
    Batch Size & 32 & Dynamic during training & 64 \\
    \hline
    Learning Rate & Start: 0.0051 & Policy: 3e-6, Discount: 1e-3 & 0.01 \\
    \hline
    Optimizer & Adam & Adam & Adam \\
    \hline
    Loss Function & \texttt{nn.BCE loss} & BCE loss & \texttt{nn.BCE loss} \\
    \hline
    Use GPU & Yes & Yes & Device dependent \\
    \hline
    Model Specific Parameters & RNN, Decoder dimensions, etc. & Policy, Discriminator settings & Model architecture details (e.g., layer sizes) \\
    \hline
  \end{tabular}
  }
  \label{tab:training_comparison}
\end{table}

\subsection{Performance Evaluation and Validation}

\subsubsection{Model Validation}
The validation phase was meticulously designed to test each model's capability to accurately impute missing data and generate synthetic datasets that reflect realistic well log characteristics. Extensive testing was conducted using a combination of real and synthesized datasets, allowing for the assessment of the models' performance under varied conditions. The testing setup included the application of mask matrices to simulate different missing data scenarios and the use of optimization algorithms to enhance the models' accuracy and reliability.

\subsubsection{Iterative Refinement and Feedback Loop}
Based on the initial results, models underwent iterative refinement processes where feedback from performance evaluations was used to adjust and optimize model parameters. This process included performance evaluation using metrics like $R^2$, MAE, MAPE, mean relative log error (MRLE), parameter adjustments (e.g., learning rate, hidden layer size), re-training, and validation. These were key components of this iterative process. The following subsections provide detailed definitions and applications of these performance metrics, as well as their roles in the iterative refinement process. This feedback loop, applied to all four workflows (TSGAN, SegGAN, BRITS, and NAOMI), was crucial in achieving the high standards required for the study, leading to progressively better outcomes in subsequent trials.

\textbf{Performance Metrics}

To evaluate the effectiveness of the models, we employed key performance metrics, which are crucial for both synthetic data generation and data imputation tasks. The metrics used include $R^2$, MAE, MRLE, and MAPE, and their definitions and applications are as follows:

\begin{itemize}
    \item \textbf{$R^2$ (Coefficient of Determination)}: This metric indicates the proportion of the variance in the dependent variable that is predictable from the independent variables. It is calculated as:
    \[
    R^2 = 1 - \frac{\sum_{i=1}^n (y_i - \hat{y}_i)^2}{\sum_{i=1}^n (y_i - \overline{y})^2} \tag{6}
    \]
    where $y_i$ are the actual values, $\hat{y}_i$ are the predicted values, and $\overline{y}$ is the mean of the actual values. A higher $R^2$ value indicates better model performance in capturing data variance. For this study, $R^2$ is computed for each log type and then averaged to get an overall $R^2$ for the dataset. $R^2$ is used in both synthetic data generation and data imputation tasks.
    
    \item \textbf{MAE (Mean Absolute Error)}: Measures the average magnitude of errors between the predicted and actual values, defined as:
    \[
    MAE = \frac{1}{n} \sum_{i=1}^n \left| y_i - \hat{y}_i \right| \tag{7}
    \]
    where $y_i$ are the actual values and $\hat{y}_i$ are the predicted values. MAE is computed for each log type and then averaged to get an overall MAE for the dataset. MAE is used in both synthetic data generation and data imputation tasks.
    
    \item \textbf{MAPE (Mean Absolute Percentage Error)}: This metric measures the average magnitude of errors between the predicted and actual values in percentage terms, defined as:
    \[
    MAPE = \frac{1}{n} \sum_{i=1}^n \left|\frac{y_i - \hat{y}_i}{y_i}\right| \times 100 \tag{8}
    \]
    where $y_i$ are the actual values and $\hat{y}_i$ are the predicted values \cite{hyndman2006another}(Hyndman \& Koehler, 2006). MAPE is computed for each log type and then averaged to get an overall MAPE for the dataset. MAPE is primarily used for data imputation tasks.
    
    \item \textbf{MRLE (Mean Relative Log Error)}: This metric emphasizes the model’s precision on a logarithmic scale, particularly useful for datasets with a wide range of values. It is calculated as:
    \[
    MRLE = \frac{1}{n} \sum_{i=1}^n \left| \log(y_i+1) - \log(\hat{y}_i+1) \right| \tag{9}
    \]
    where $y_i$ are the actual values and $\hat{y}_i$ are the predicted values \cite{armstrong1992error}. MRLE is computed for each log type and then averaged to get an overall MRLE for the dataset. MRLE is primarily used for synthetic data generation.
\end{itemize}

These metrics were crucial in the iterative refinement process, guiding the adjustments in model parameters and the optimization of model performance.

\subsubsection{Comparative Analysis}
To underscore the efficacy of this approach, the results were compared with traditional methods and other contemporary deep learning approaches. This comparative analysis revealed that the models not only filled missing data with high accuracy but also enhanced the resolution and quality of well log data, supporting more robust geological interpretations.

\subsubsection{Visual Inspection, Case Studies, Dimensionality Reduction, and Performance Metrics}
In this study, visual inspections of imputed logs alongside detailed case studies of specific wells were imperative for demonstrating the practical application and effectiveness of the models. These case studies provided demonstrations of how the approach successfully corrected substantial data gaps, offering insights that were previously obscured by missing data. To further aid in the intuitive understanding of the models' efficacy, visualization and dimensionality reduction techniques were employed. Techniques such as t-SNE and PCA were crucial in visualizing the high-dimensional generated data. These methods allowed for a better understanding of the complex interactions within the data and highlighted the realistic imputation performed by the models. By using these visualization techniques, not only was the accuracy of the imputations visually confirmed in specific case studies, but the broader applicability and reliability of the methods were also efficiently demonstrated across various well log parameters. The integrated use of case studies and advanced visualization techniques enriched the analysis, making the technical outcomes more accessible and easier to interpret for both technical and non-technical stakeholders.

Model performance was quantitatively assessed using $R^2$, MAE, MRLE, and MAPE across different sections of the well logs to ensure a thorough evaluation of the imputation accuracy. Each methodological step, from data acquisition through model validation, was designed not just to fill technical gaps but to advance the understanding of oil and gas reservoir characteristics.

\section{Results and Discussion}

\subsection{Synthetic data generation with TSGAN for 1D well logging}
This section presents the results of employing TSGAN to generate synthetic well logging data, following the detailed methodology outlined in Section 2.1.
The focus was on generating four types of logs: GR, DT, RHOB, and NPHI, as well as derived quantities like porosity (PHI), effective porosity (PHIECAL), and volumetric shale (VSHALE). These logs were prepared through rigorous data preprocessing (Phase 1), which included data cleaning, normalization, and segmentation to ensure high-quality input data. Volumetric shale and porosity were derived as follows:
\[
V_{\text{SHALE}} = \frac{GR - GR_{\text{min}}}{GR_{\text{max}} - GR_{\text{min}}} \tag{10}
\]
and PHI through density porosity calculations:
\[
\Phi_d = \frac{RHOB_{\text{matrix}} - RHOB_{\text{bulk}}}{RHOB_{\text{matrix}} - RHOB_{\text{fluid}}} \tag{11}
\]

\subsubsection{Quantitative performance evaluation}
The detailed performance results are summarized in Table 3, which provides a comparative analysis of the $R^2$, MAE, and MRLE values for both real and synthetic datasets. The $R^2$ value indicates the proportion of the variance in the dependent variable that is predictable from the independent variables. The MAE measures the average magnitude of errors between the predicted and actual values, and MRLE emphasizes the model’s precision on a logarithmic scale.

\begin{table}[h]
\centering
\caption{Performance metrics for real vs. synthetic data evaluation}
\begin{tabular}{|l|l|l|l|}
\hline
\textbf{Data Type} & \textbf{R²} & \textbf{MAE} & \textbf{MRLE} \\ \hline
Real      & 0.894544 & 0.033437 & 0.001398  \\ \hline
Synthetic & 0.923233 & 0.034736 & 0.001194  \\ \hline
\end{tabular}
\end{table}

The results indicate that the synthetic data closely matches the real data in terms of variability and structure, as evidenced by a higher $R^2$ value for the synthetic data. Although there is a slight increase in MAE, this metric quantifies the average deviation between the synthetic data and the real data values. The lower MRLE indicates higher precision on a logarithmic scale, which is crucial for maintaining data consistency in geological applications. This outcome reflects the effectiveness of the data preprocessing (Phase 1) and model architecture (Phase 2), as shown in Figure 2.

\subsubsection{Visual analysis}
To verify the diversity and authenticity of the synthetic data generated by TSGAN, dimensionality reduction techniques such as PCA and t-SNE were employed to illustrate the distribution of real and synthetic data, affirming the synthetic data’s fidelity to realistic geological patterns. These visual tools helped substantiate the statistical results, providing a dual confirmation of the model’s effectiveness.

Figure 8 provides a qualitative comparison of the real and synthetic data distributions. The PCA results, which capture 85\% of the total variance with the first two principal components, depict how closely the synthetic data (orange points) aligns with the real data (blue points), demonstrating the model’s effectiveness in capturing the true data distribution.
\begin{figure}[htbp]
    \centering
    \begin{minipage}{1\textwidth}
        \centering
        \includegraphics[width=\textwidth]{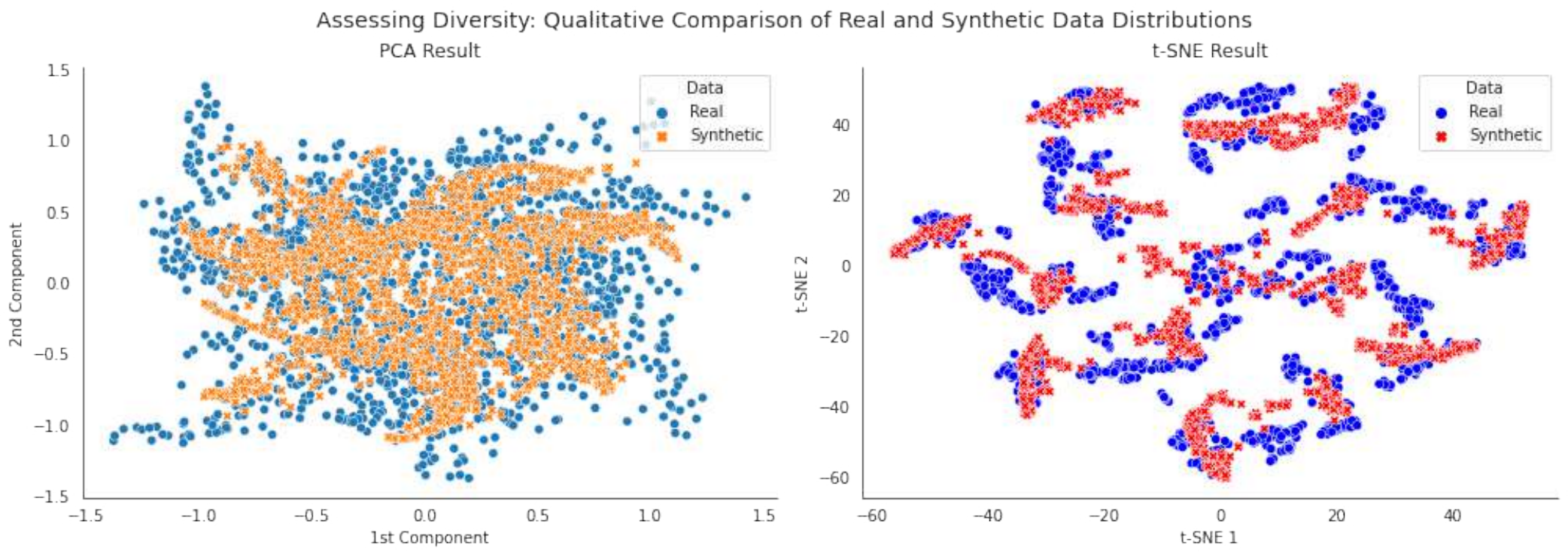}  % Image for (a) and (b)
        \subcaption{PCA analysis: This plot shows how closely the synthetic data (orange points) align with the real data (blue points), demonstrating the effectiveness of TSGAN in capturing the true data distribution. The first two principal components capture 85\% of the total variance.}
        \vskip\baselineskip
        \subcaption{t-SNE Analysis: The t-SNE plot displays the distribution of real and synthetic data, highlighting the model’s ability to maintain intrinsic properties of different data types.}
    \end{minipage}
    \vskip\baselineskip
    \begin{minipage}{1\textwidth}
        \centering
        \includegraphics[width=\textwidth]{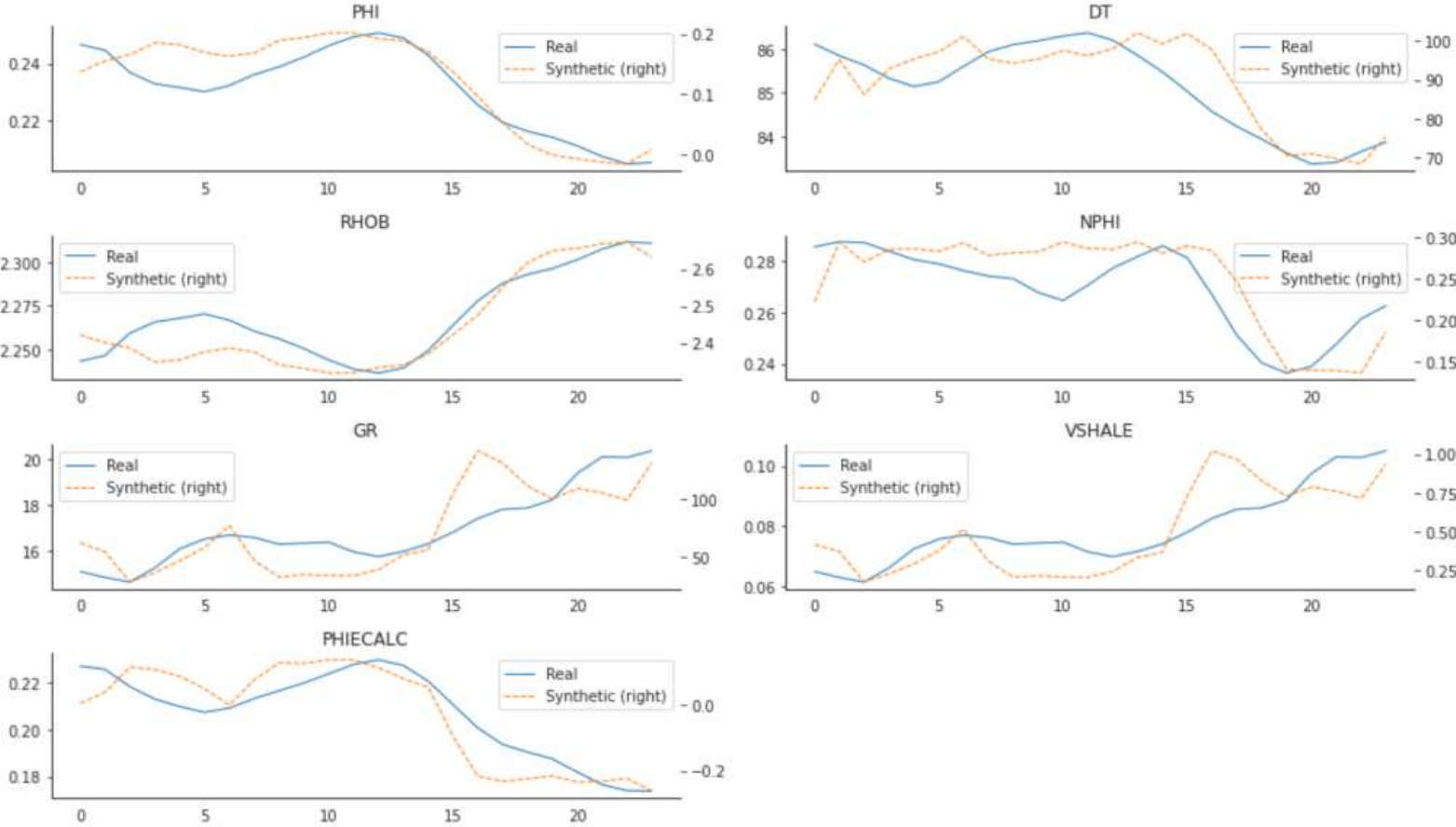}  % Image for (c)
        \subcaption{Real vs. synthetic data depth curves: Comparison of real and synthetic well log data across various parameters (PHI, DT, RHOB, NPHI, GR, VSH, and PHIECAL) with depth. The x-axis represents depth in meters, and the y-axis represents the respective log values.}
        \label{3c}
    \end{minipage}
    \caption{Visualization of synthetic and real data distributions.}
    \label{4}
\end{figure}

\subsubsection{Discussion on model effectiveness}

The TSGAN model has proven effective in replicating the necessary characteristics of well logs. This capability is key for enhancing subsurface data analysis, especially in scenarios where actual data may be sparse or incomplete. The high degree of statistical and visual alignment with real data underscores the model's potential for broader applications in geological research.

To quantify the alignment between the real and synthetic data distributions, we conducted the Kolmogorov-Smirnov (KS) test, calculated Pearson correlation coefficients (PCC), and computed the Kullback-Leibler (KL) divergence. These metrics provide a robust statistical comparison of the distributions, validating the model’s effectiveness.

\begin{table}[htbp]
\centering
\caption{Statistical comparison of real and synthetic data}
\begin{tabular}{|l|l|l|l|l|}
\hline
\textbf{Log Type} & \textbf{KS statistic} & \textbf{p-value} & \textbf{PCC} & \textbf{KL divergence} \\
\hline
PHI & 0.03 & 0.57 & 0.93 & 0.025 \\
DT & 0.05 & 0.42 & 0.91 & 0.023 \\
RHOB & 0.04 & 0.53 & 0.92 & 0.027 \\
NPHI & 0.02 & 0.65 & 0.94 & 0.021 \\
GR & 0.03 & 0.59 & 0.90 & 0.026 \\
VSHALE & 0.04 & 0.50 & 0.91 & 0.024 \\
\hline
\end{tabular}
\end{table}

Table 4 indicates that the synthetic data closely matches the real data in terms of variability and structure. The KS test results show no significant difference between the distributions of real and synthetic data \cite{massey1951kolmogorov}. The PCCs demonstrate a high degree of linear correlation between the real and synthetic data \cite{benesty2009pearson}. The low KL divergence values indicate a close match between the real and synthetic data distributions \cite{kullback1951information}.

These statistical comparisons, along with the visual alignment observed in the PCA and t-SNE analyses, confirm the model’s effectiveness in generating high-fidelity synthetic well log data.

\subsubsection{Significance in geosciences}
The application of TSGAN to well log data extends the tool's utility to the geosciences, a field where enhanced synthetic data can significantly improve subsurface understanding and interpretation. This adaptation is particularly vital in regions where well log data are sparse or incomplete, offering a method to enrich datasets without the extensive costs and challenges of new data acquisition. By addressing the unique requirements of high-dimensional well log data, this study not only pioneers TSGAN's use in this field but also establishes a foundational approach for future synthetic data generation in geosciences.

\subsubsection{Future directions}
This study confirms the potential of TSGAN to generate high-fidelity synthetic well log data. Future work will focus on expanding the range of log types and integrating variable geological conditions to further test the model’s robustness and applicability. Additionally, exploring the use of TSGAN in conjunction with other advanced ML techniques could provide new insights into optimizing well log data analysis and improving predictive models in geosciences.

\subsection{Enhanced imputation of missing values in 1D well logging data}

The challenge of missing values in well logging data is a significant hurdle in geophysical analysis and decision-making. Efficient and accurate imputation of these missing values is critical for reliable geological interpretations and subsequent exploratory actions. This section delves into the performance of three advanced machine learning models—GAN, NAOMI, and BRITS—in imputing missing well log data, providing a detailed comparison of their effectiveness across various sections of the logs.

\subsubsection{Comparative performance metrics}
Table 5 explores comparative performance metrics of different models (NAOMI, GAN, BRITS) across selected sections using MAPE, MAE, and R² values. The results demonstrate the effectiveness of each model in terms of imputation accuracy for the specific sections of well log data. For this study, sections of the DT log were artificially corrupted with missing values to simulate realistic scenarios. The missing values were randomly distributed across the sections, each consisting of 5, 10, and 50 data points respectively, as illustrated in Figure 9. The test set consisted of these corrupted sections, and the performance metrics were calculated by comparing the imputed values to the original values.

\begin{table}[htbp]
\centering
\caption{Imputation performance comparison}
\begin{tabular}{|l|l|l|l|l|}
\hline
\textbf{Model} & \textbf{Section (Data points)} & \textbf{MAPE} & \textbf{MAE} & \textbf{R²} \\
\hline
\multirow{3}{*}{NAOMI} & 5 & 2.291 & 0.004 & 0.236 \\
 & 10 & 2.355 & 0.004 & 0.888 \\
 & 50 & 151.154 & 0.032 & 0.594 \\
\hline
\multirow{3}{*}{SeqGAN (Proposed)} & 5 & 1.109 & 0.002 & \textbf {0.866} \\
 & 10 & 8.320 & 0.012 & \textbf{0.921} \\
 & 50 & 166.63 & 0.030 & 0.574 \\
\hline
\multirow{3}{*}{BRITS} & 5 & 0.004 & 0.004 & 0.438 \\
 & 10 & 0.005 & 0.005 & 0.899 \\
 & 50 & 0.049 & 0.049 & 0.12 \\
\hline
\end{tabular}
\label{tab:imputation_comparison}
\end{table}

The results indicate:

\begin{itemize}
    \item \textbf{Shorter Sections (Length 5 data points):} The GAN model exhibits superior performance with the lowest MAPE and MAE values alongside a high R², indicating a strong adherence to the true data variability.
    \item \textbf{Longer Sections (Length 10 data points):} GAN again shows the best fit with an impressive R² of 0.921, suggesting robust predictive power and reliability in handling complex data structures.
    \item \textbf{Extended Sections (Length 50 data points):} The NAOMI model scored a MAPE of 151.154, MAE of 0.032, and an R² of 0.594, showing its ability to manage larger sections but with less accuracy compared to shorter sections. In contrast, GAN's performance slightly decreased with MAPE of 166.63, MAE of 0.030, and R² of 0.574. BRITS performed significantly lower in this section with MAPE of 0.049, MAE of 0.049, and R² of 0.12.
\end{itemize}

These findings underscore the capability of advanced generative models to effectively impute missing well log data, thereby enhancing data quality and supporting robust subsurface geological analysis.

\subsubsection{Model performance analysis}
Figure 9 illustrates the performance of the GAN, NAOMI, and BRITS models in imputing missing DT log data. This side-by-side comparison helps in visualizing the strengths and weaknesses of each model:

\begin{enumerate}
    \item \textbf{Baseline Imputation:}
    \begin{itemize}
        \item \textit{Continuous Prediction Accuracy (Figure 9a):} Provides continuous prediction accuracy without significant data gaps, offering a reference point for comparison.
    \end{itemize}
    \item \textbf{GAN Model:}
    \begin{itemize}
        \item \textit{Continuous Prediction Accuracy (Figure 9b):} The GAN model maintains high fidelity across standard segments without significant data gaps, showcasing its ability to predict continuous data accurately.
        \item \textit{Highlighted Imputation (Figure 9b):} When tested against a significant missing data segment, the GAN model effectively handles imputation, with the imputed values highlighted in orange, demonstrating its interpolation strengths.
    \end{itemize}
    \item \textbf{NAOMI Model:}
    \begin{itemize}
        \item \textit{Effective Reconstruction (Figure 9c):} The NAOMI model shows robust performance in managing larger sections with complex geological features, effectively reconstructing missing values (highlighted in orange).
    \end{itemize}
    \item \textbf{BRITS Model:}
    \begin{itemize}
        \item \textit{Adequate Imputation (Figure 9d):} The BRITS model highlights reasonable accuracy but less precision compared to GAN and NAOMI, particularly in specific contexts with missing values (highlighted in orange).
    \end{itemize}
\end{enumerate}

\begin{figure}[htbp]
    \centering
    \begin{minipage}{1\textwidth}
        \centering
        \includegraphics[width=\textwidth]{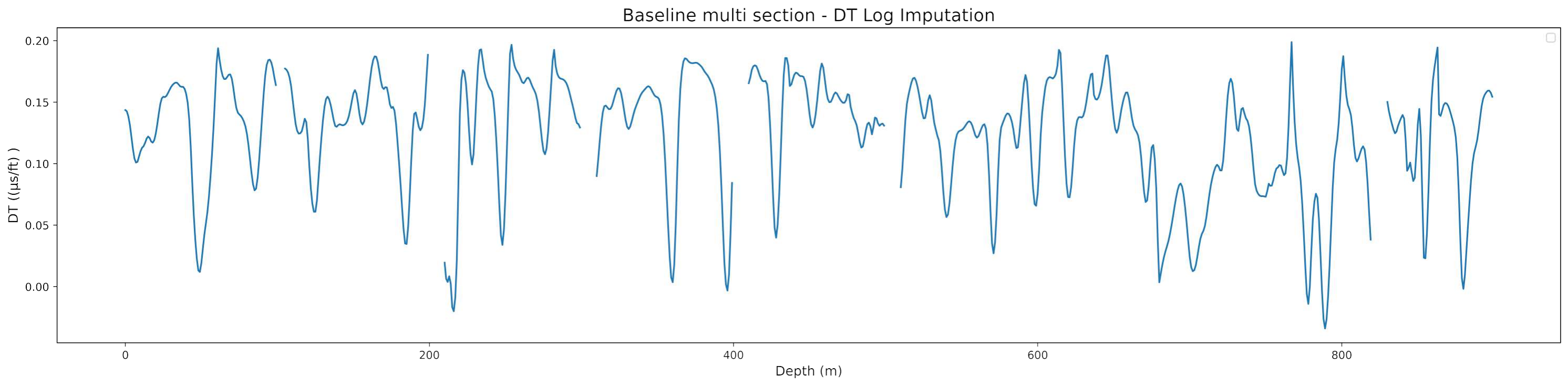}  % Image for (a)
        \subcaption{Baseline imputation with no highlighted gaps, showcasing continuous prediction accuracy.}
    \end{minipage}\hfill
    \begin{minipage}{1\textwidth}
        \centering
        \includegraphics[width=\textwidth]{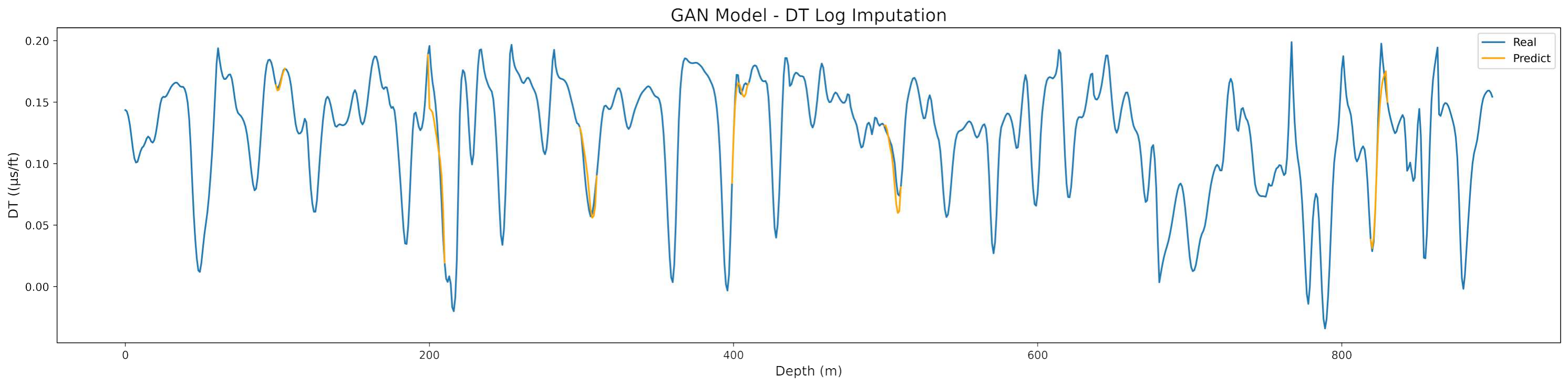}  % Image for (b)
        \subcaption{GAN Model - Highlighted imputation of missing data (orange segment).}
    \end{minipage}\hfill
    \begin{minipage}{1\textwidth}
        \centering
        \includegraphics[width=\textwidth]{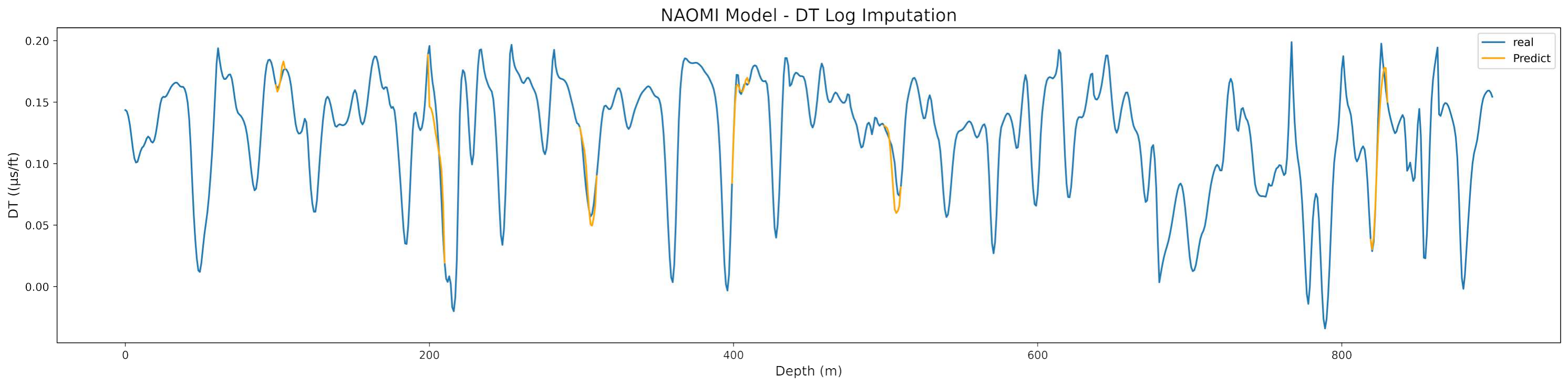}  % Image for (c)
        \subcaption{NAOMI Model - Effective reconstruction of missing values (orange segment).}
    \end{minipage}\hfill
    \begin{minipage}{1\textwidth}
        \centering
        \includegraphics[width=\textwidth]{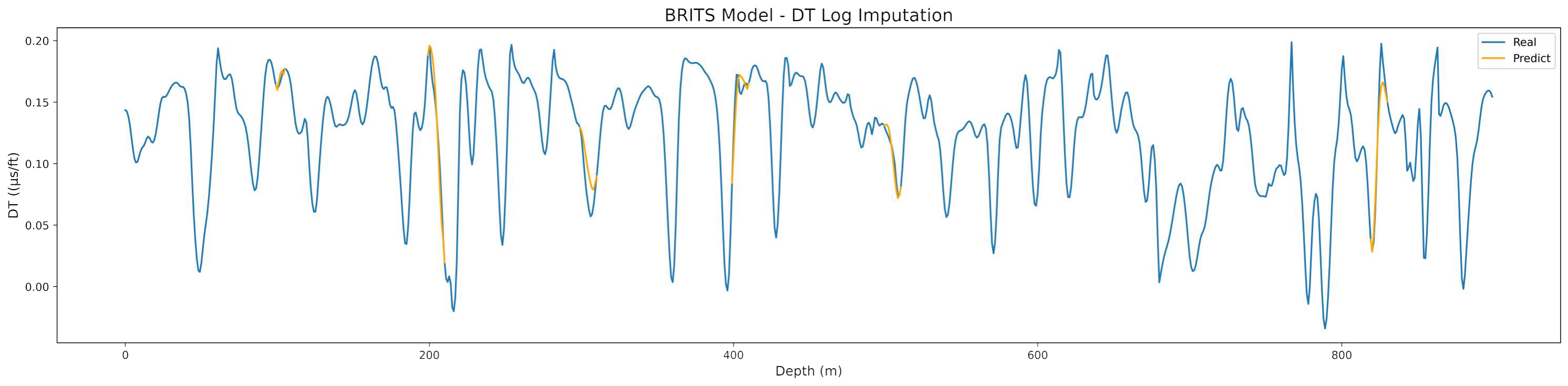}  % Image for (d)
        \subcaption{BRITS Model - Adequate but less precise imputation of missing values (orange segment).}
    \end{minipage}
    \caption{Comparison of model performances in DT log imputation for short sections (5 and 10 data points).}
    \label{fig:DT_imputation_comparison}
\end{figure}

The study further explores the imputation performance of GAN, NAOMI, and BRITS models across various section lengths of well log data. New results have been incorporated to extend the analysis to broader scenarios. These findings are particularly revealing in terms of each model's capacity to manage larger gaps in data, as shown in Figure 10. Each subplot (Figures 10a, 10b, 10c, 10d) illustrates the nuanced performance differences between the models, providing a clear visual representation of their capabilities across different geological layers and conditions. This side-by-side comparison helps in visualizing the strengths and weaknesses of each model in handling more extensive missing data segments:
\begin{itemize}
    \item \textbf{Baseline imputation:}
    \begin{itemize}
        \item Continuous prediction accuracy (Figure 10a): Provides continuous prediction accuracy without significant data gaps, offering a reference point for comparison.
    \end{itemize}
    \item \textbf{GAN model:}
    \begin{itemize}
        \item Robust performance (Figure 10c): Demonstrates reliability across varied conditions, maintaining high accuracy even in scenarios with extensive missing data. The imputed values are highlighted in orange.
    \end{itemize}
    \item \textbf{NAOMI model:}
    \begin{itemize}
        \item Enhanced capability (Figure 10b): Shows effectiveness in handling larger data sections with intricate geological features, indicating its robustness in extensive imputation tasks. The imputed values are highlighted in orange.
    \end{itemize}
    \item \textbf{BRITS model:}
    \begin{itemize}
        \item Adequate imputation (Figure 10d): Highlights reasonable accuracy but less precision compared to GAN and NAOMI, particularly in contexts with extensive missing values. The imputed values are highlighted in orange.
    \end{itemize}
\end{itemize}

\begin{figure}[htbp]
    \centering
    \begin{minipage}{1\textwidth}
        \centering
        \includegraphics[width=\textwidth]{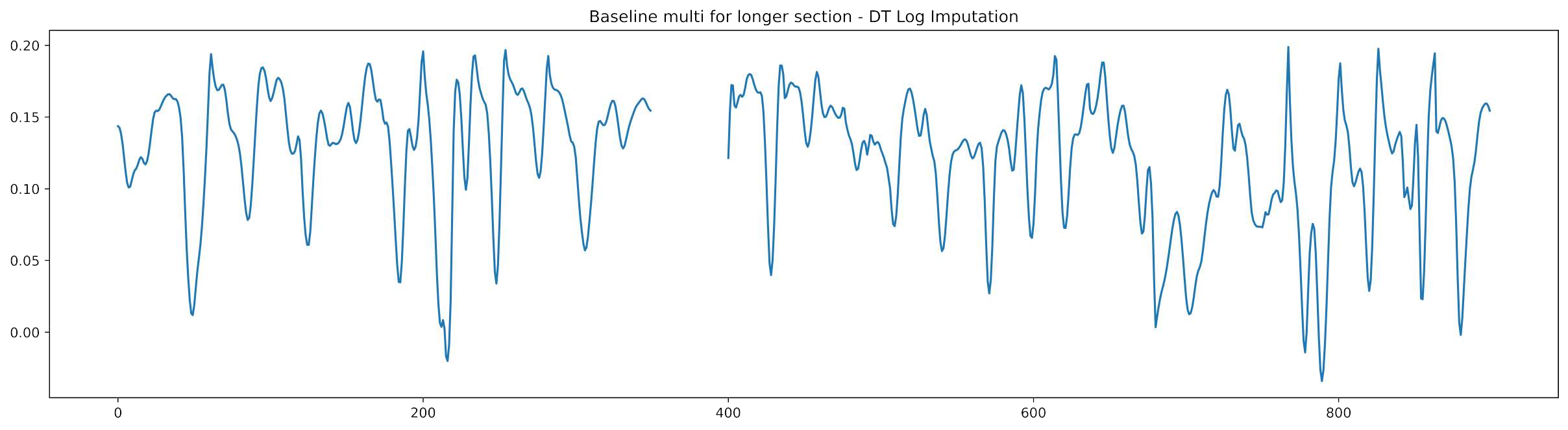}  % Image for (a)
        \subcaption{Baseline imputation with no highlighted gaps, showcasing continuous prediction accuracy.}
    \end{minipage}
    
    \begin{minipage}{1\textwidth}
        \centering
        \includegraphics[width=\textwidth]{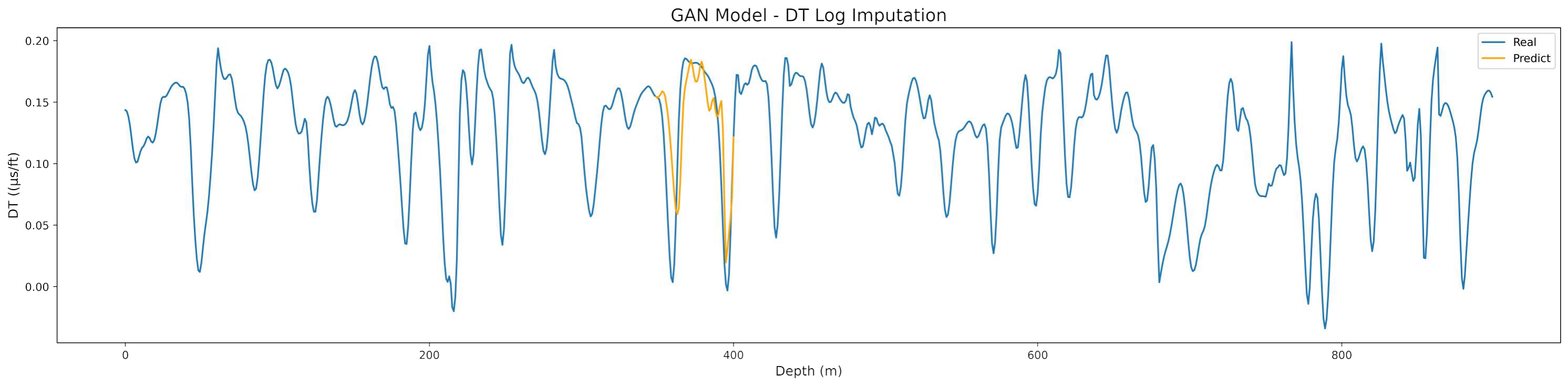}  % Image for (b)
        \subcaption{GAN model performance: Demonstrates robustness across varied conditions, maintaining high accuracy and reliability, even in extensive missing data scenarios.}
    \end{minipage}
    
    \begin{minipage}{1\textwidth}
        \centering
        \includegraphics[width=\textwidth]{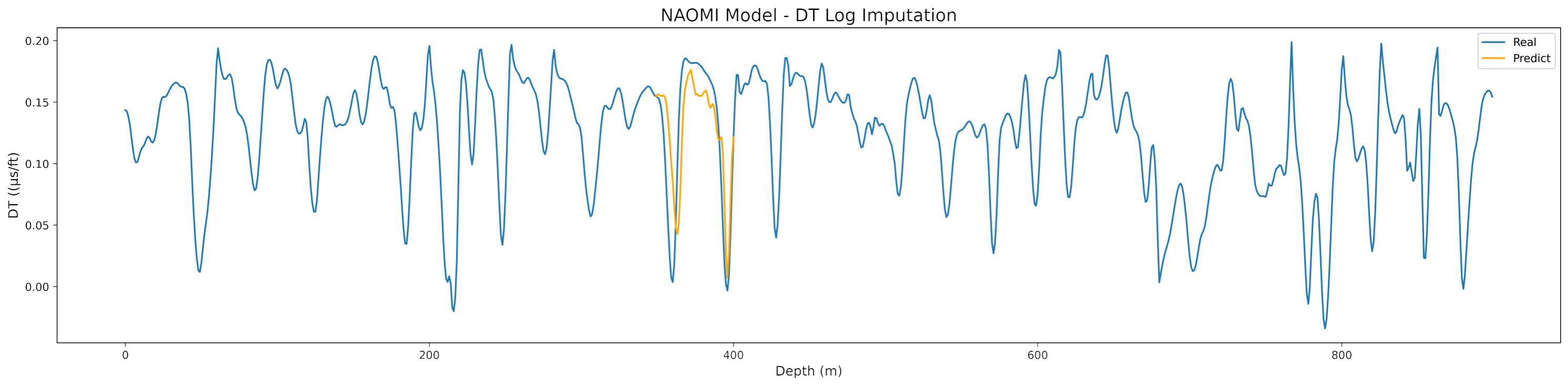}  % Image for (c)
        \subcaption{NAOMI model performance: Shows enhanced capability in segment 50, indicating its effectiveness in handling larger data sections with complex geological features.}
    \end{minipage}
    
    \begin{minipage}{1\textwidth}
        \centering
        \includegraphics[width=\textwidth]{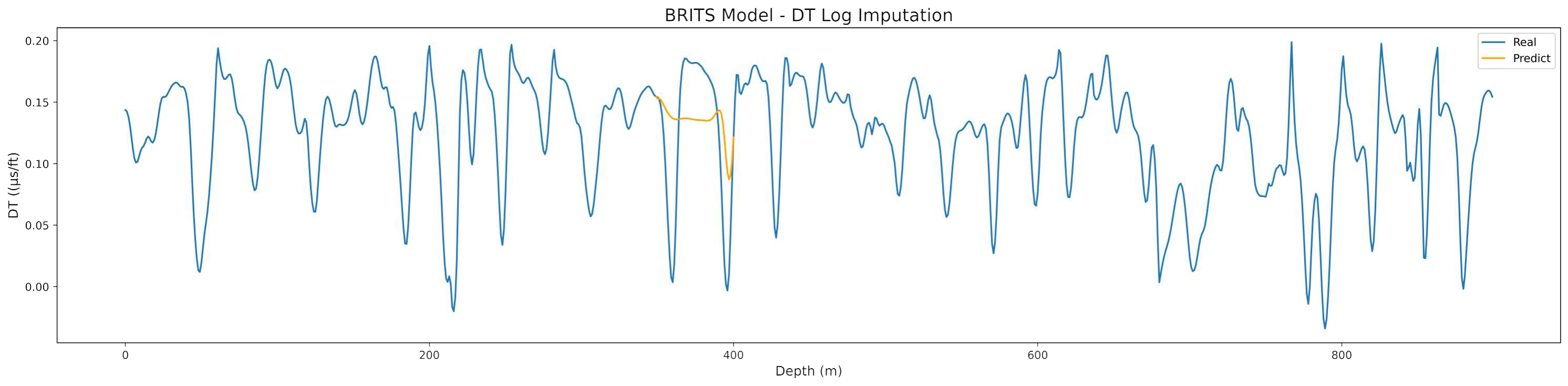}  % Image for (d)
        \subcaption{BRITS model performance: Adequate but less precise imputation of missing values.}
    \end{minipage}
    
    \caption{Comparison of performance of imputation models for longer sections (50 data points).}
    \label{10}
\end{figure}

\subsubsection{Summary of comparative analysis}

The comprehensive evaluation of imputation techniques, as elucidated in the accompanying Table 5 and illustrated in Figures 9 and 10, provides an in-depth comparison of each model's effectiveness in handling missing well log data.

\begin{itemize}
    \item \textbf{GAN model}: Notably stands out for its exceptional performance in shorter sections, demonstrating a robust ability to adhere closely to the actual data patterns with minimal error. This model's adeptness at managing smaller segments makes it particularly valuable for scenarios where high precision is necessary over short intervals.
    
    \item \textbf{NAOMI model}: While the GAN model shows impressive results in shorter sections, the NAOMI model proves to be highly effective across various geological contexts, maintaining consistent performance even in extensive missing data scenarios. Its capability to interpolate missing values seamlessly, regardless of the data complexity, highlights its adaptability and reliability.
    
    \item \textbf{BRITS model}: Demonstrates proficiency in reconstructing missing data sequences with reasonable accuracy. However, its performance is adequate but not exceptional, indicating the need for further refinement to match the precision observed with the GAN and NAOMI models in different contexts.
\end{itemize}

This side-by-side performance comparison serves as a critical guide for practitioners, aiding in the selection of the most appropriate imputation model based on the specific conditions and demands of the dataset. By aligning model strengths with project requirements, this analysis ensures that the chosen method maximizes data recovery and enhances the reliability of subsurface interpretations.

\section{Conclusion}

This study presents a significant advancement in applying generative adversarial networks to enhance the integrity and usability of well log data. Implementing a dual-framework of TSGAN and SeqGAN, we demonstrated substantial improvements over traditional data imputation and synthetic data generation methods. TSGAN excels in generating high-quality synthetic data that faithfully replicates real-world geological attributes, while SeqGAN provides precise imputation of missing data sequences. Comparative analyses with BRITS and NAOMI underscored the superiority of our methods across various test scenarios and geological contexts, offering robust capabilities for geoscientists and engineers in oil and gas exploration decision-making.

SeqGAN's superior performance leverages adversarial training to capture sequential dependencies inherent in well log data more effectively. Unlike traditional methods, SeqGAN uses contextual information from adjacent data points to predict missing values, ensuring coherent and realistic imputations. This capability preserves spatial coherence necessary for accurate geological interpretations and synthetic dataset fidelity, validated through principal component analysis (PCA) and t-distributed stochastic neighbor embedding (t-SNE). These findings highlight GANs' promising applications in solving geological challenges and reducing uncertainties in reservoir characterization, thereby enhancing predictive reliability in exploration processes.

The integration of two specialized sequence-based GANs sets a new benchmark for data integrity and utility in geosciences. This approach not only advances GANs' role in geophysical data analysis but also suggests future innovations that could redefine industry standards for analytical precision and resource exploration effectiveness. Continued advancements in this area promise to enhance subsurface modeling capabilities, optimizing exploration strategies and resource recovery efficiency.
\section*{CRediT Authorship Contribution Statement}
Abdulrahman Al-Fakih: Formal analysis, Methodology, Software, Writing – original draft, Data preparation, Code creation. 
A. Koeshidayatullah: Resources, Supervision, Review \& editing. 
Tapan Mukerji: Conceptualization, Review \& editing, Scientific additions. 
Sadam Al-Azani: Review \& editing, Re-enhancing figures. 
SanLinn I. Kaka: Supervision, Review \& editing.

\section*{Declaration of Competing Interest and Use of Generative AI}
The authors affirm that they have no known competing financial interests or personal relationships that may have influenced the work presented in this paper. 
During the preparation of this work, the author(s) used the ChatGPT language model from OpenAI to refine grammar and enhance text coherence in this article. After using this tool/service, the author(s) reviewed and edited the content as needed and take(s) full responsibility for the content of the publication.

\section*{Data Availability}
The datasets used and/or analyzed in this study are available upon reasonable request from the corresponding author, Abdulrahman Al-Fakih, via email at \texttt{alja2014ser@gmail.com}.

\section*{Acknowledgements}
The authors would like to extend their sincere gratitude to the College of Petroleum Engineering at King Fahd University of Petroleum and Minerals (KFUPM) for their invaluable support in facilitating the presentation of this work at international conferences. We also acknowledge the NLOG website and Utrecht University for generously providing the dataset used in this study. Additionally, we wish to express our appreciation to the SDAIA-KFUPM-JRC-AI Research Center for their technical assistance.


%%% Comment out this section when you \begin{thebibliography}{149}

% Now the Appendix section comes after the references
\appendix
\section{Outlier detection and data cleaning techniques in well log}

% Reset figure counter and customize the format for figures in the appendix
\renewcommand{\thefigure}{\arabic{figure}A} % This will make figures numbered 1A, 2A, etc.

\begin{figure}[htbp]
    \centering
    \includegraphics[width=1\textwidth]{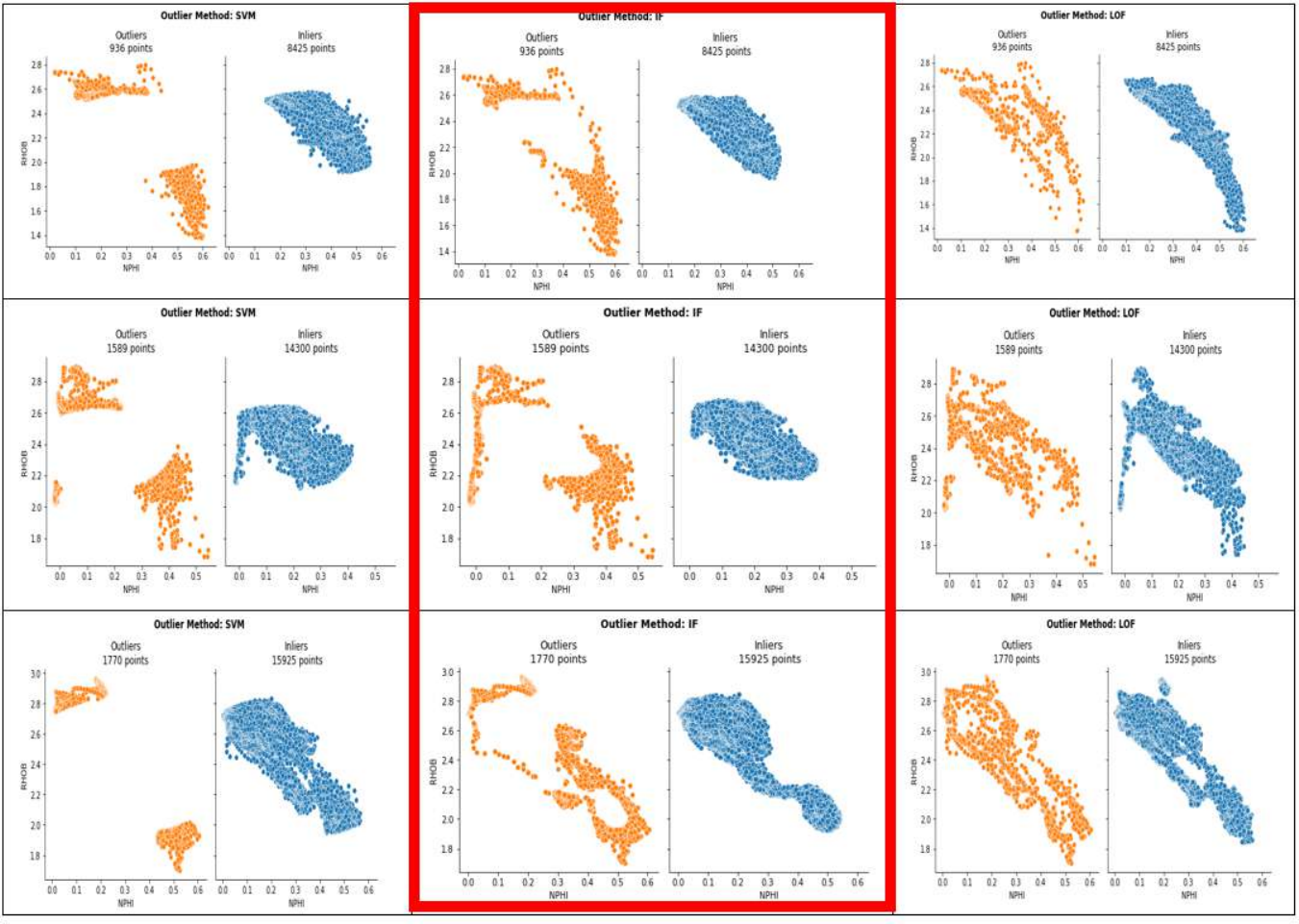}  % Use the correct file path here
    \caption{Outlier detection and removal techniques applied to NPHI and RHOB well log data using three methods: SVM, IF, and LOF.}
    \label{ure1A}
\end{figure}

\begin{figure}[htbp]
    \centering
    \includegraphics[width=1\textwidth]{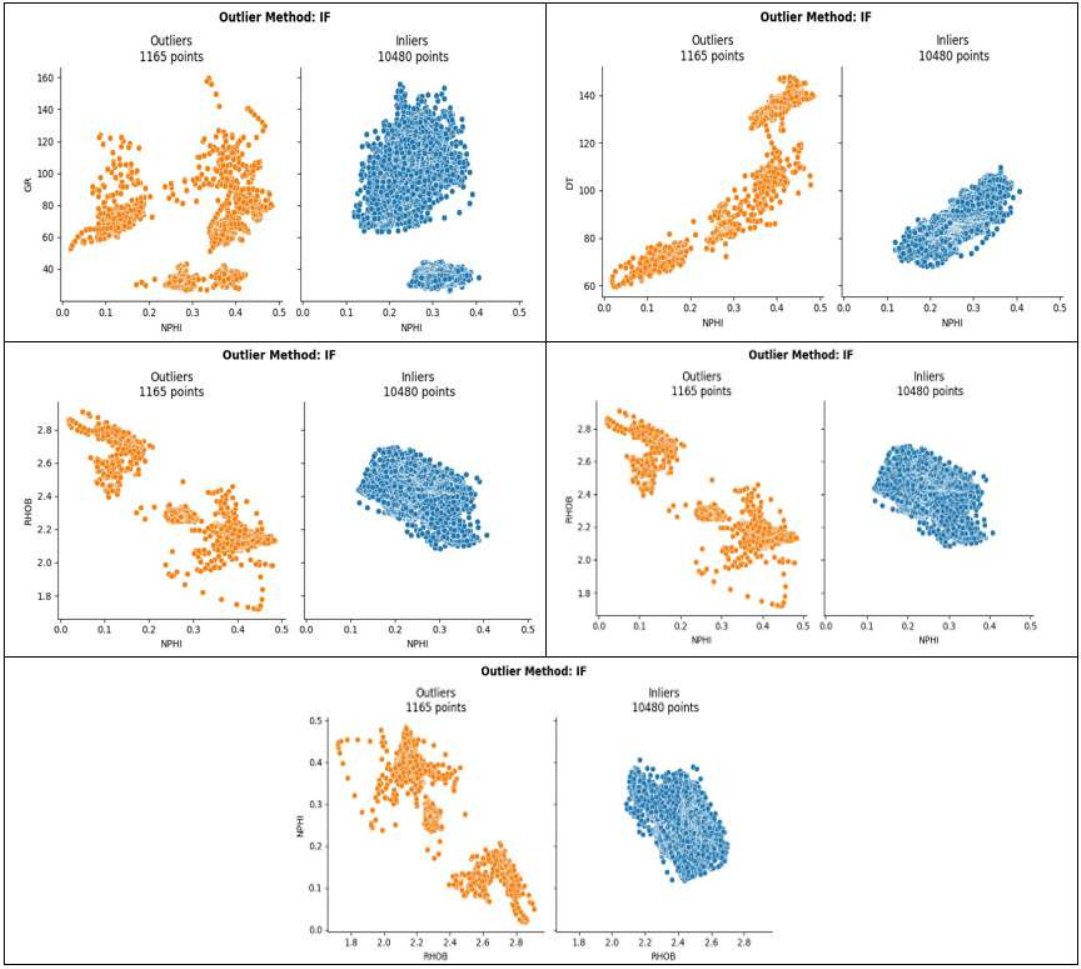}  % Use the correct file path here
    \caption{Outlier detection techniques applied to all well logs (GR, DT, RHOB, NPHI, ILD) for well F05-02 using the IF method. The figure shows the identification and removal of approximately 10\% of the data points as outliers (orange points) across all logs, with the inliers (blue points) being retained for subsequent analysis.}
    \label{ure2A}
\end{figure}

\begin{figure}[htbp]
    \centering
    \includegraphics[width=1\textwidth]{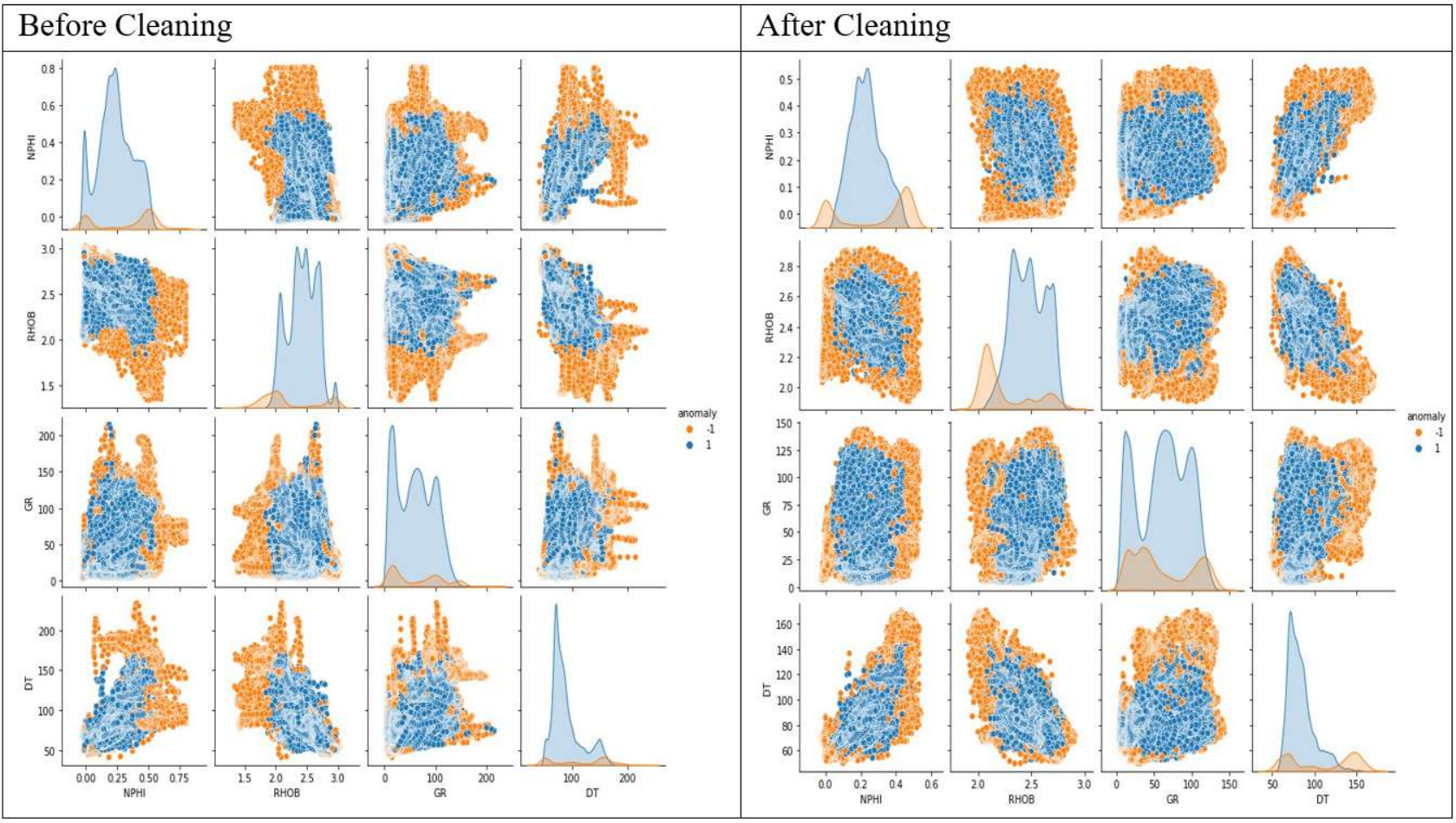}  % Use the correct file path here
    \caption{Visual comparison of well log data before and after cleaning using the isolation forest method. The removal of outliers enhances data accuracy and model reliability for reservoir property predictions.}
    \label{ure3A}
\end{figure}

\end{document}